\documentclass[aps,prd,twocolumn,superscriptaddress,showpacs,nofootinbib]{revtex4-1}

\usepackage{amsmath}
\usepackage{graphicx}
\usepackage{epstopdf}
\usepackage{latexsym}
\usepackage{amssymb}
\usepackage{hyperref}
\usepackage{comment}
\hypersetup{
    colorlinks=true,
    linkcolor=blue,
    citecolor=blue,
    urlcolor=blue
}
\usepackage{array}
\usepackage{enumitem}
\usepackage{float}
\usepackage{lipsum}  

\begin{document}


\title{Searching for dark matter -- dark energy interactions: going beyond the conformal case}

\author{Carsten van de Bruck}
\email{c.vandebruck@sheffield.ac.uk}
\affiliation{Consortium for Fundamental Physics, School of Mathematics and Statistics, University of Sheffield, Hounsfield Road, Sheffield S3 7RH, UK} 
\author{Jurgen Mifsud}
\email{jmifsud1@sheffield.ac.uk}
\affiliation{Consortium for Fundamental Physics, School of Mathematics and Statistics, University of Sheffield, Hounsfield Road, Sheffield S3 7RH, UK}  

\date{\today}

\begin{abstract}
We consider a generic cosmological model which allows for non--gravitational direct couplings between dark matter and dark energy. The distinguishing cosmological features of these couplings can be probed by current cosmological observations, thus enabling us to place constraints on this generic interaction which is composed of the conformal and disformal coupling functions.
We perform a global analysis in order to independently constrain the conformal, disformal, and mixed interactions between dark matter and dark energy by combining current data from: \textit{Planck} observations of the cosmic microwave background radiation anisotropies, a combination of measurements of baryon acoustic oscillations, a supernovae Type Ia sample, a compilation of Hubble parameter measurements estimated from the cosmic chronometers approach, direct measurements of the expansion rate of the Universe today, and a compilation of growth of structure measurements. We find that in these coupled dark energy models, the influence of the local value of the Hubble constant does not significantly alter the inferred constraints when we consider joint analyses that include all cosmological probes. Moreover, the parameter constraints are remarkably improved with the inclusion of the growth of structure data set measurements. We find no compelling evidence for an interaction within the dark sector of the Universe. 
\end{abstract}


\maketitle



\section{Introduction}
\label{sec:introduction}
The rapid progression of precision cosmology has undoubtedly lead to a wide spectrum of cosmological probes that are able to survey different epochs of the cosmic history of our Universe. Intriguingly, the simplest cosmological framework of the concordance $\Lambda$CDM cosmology has always been found to be in an excellent agreement with these cosmological observations, and its parameters have now been determined to an impressive accuracy \cite{Ade:2015xua}. Thus, given that this model has survived this avalanche of high precision data, robust constraints on new physics beyond the $\Lambda$CDM model are always getting tighter \cite{Ade:2015rim,Ade:2015lrj,Zhao:2008bn,Raveri:2014cka,Hu:2015rva,Bellini:2015xja,Samushia:2013yga}. Nevertheless, there have been indications in the data that are not well described by the $\Lambda$CDM model. For instance, a discrepancy between the $Planck$ cosmic microwave background (CMB) radiation inferred constraint on the linear theory rms fluctuation in total matter in $8\,h^{-1}\,\mathrm{Mpc}$ spheres and the more direct measurements of the large scale structure \cite{Vikhlinin:2008ym,Ade:2013lmv,Hildebrandt:2016iqg,Heymans:2013fya}, together with a tension between the relatively high local value of the Hubble constant \cite{Riess:2016jrr} and that derived from the CMB data, have been reported. These anomalies could potentially be an indication of new phenomena which the $\Lambda$CDM model does not take into account \cite{Lesgourgues:2015wza,Bull:2015stt,Hamaguchi:2017ihw,Amendola:2016saw}, or could be caused by systematics in astrophysical data \cite{Wu:2017fpr,Bernal:2016gxb,Efstathiou:2017rgv}.   

From a theoretical perspective, despite the simplicity of the interpretation of dark energy as a cosmological constant, the concordance $\Lambda$CDM model suffers from the well--known theoretical issues of the fine--tuning and coincidence problems \cite{Weinberg:1988cp,Zlatev:1998tr}. In light of our incomplete understanding of the cosmic evolution of our Universe, the consideration of alternative models of dark energy is a rational step towards a more comprehensive view of our Universe. Consequently, a plethora of alternative dynamical dark energy models have been proposed and well investigated (see for instance Refs. \cite{Copeland:2006wr,Bolotin:2013jpa}). Since the dark sector of the Universe, composed of dark matter and dark energy, has only been indirectly observed with cosmological observations, interactions between dark matter particles beyond the gravitational force and mediated by dark energy cannot be excluded a priori \cite{Wetterich:1994bg,Holden:1999hm,Carroll:1998zi,Damour:1990tw,Gubser:2004du,Farrar:2003uw,Carroll:2008ub}.

Here we focus on a generic coupled dark energy model \cite{Mifsud:2017fsy,Jack}, in which dark matter and dark energy are interacting with one another, whereas the standard model particles follow their standard cosmological evolution (for other coupled models see Refs. \cite{Wang:2016lxa,Dutta:2017wfd,Santos:2017bqm,Yang:2017yme,Sharov:2017iue} and references therein). Thus, this model evades the tight constraints inferred from the equivalence principle and solar system tests \cite{Bertotti:2003rm,Will}. To be concrete, we assume two coupling functions in our work, these being the conformal and disformal couplings, which we will formally define in section \ref{sec:model}. In these coupled dark energy scenarios, the additional fifth--force within the dark sector of the Universe modifies the background evolution, as well as the evolution of cosmological perturbations. Coupled dark energy models with a conformal coupling function have been exhaustively explored and tight constraints have been placed on the model parameters \cite{Amendola:2003eq,Bean:2008ac,Xia:2009zzb,Amendola:2011ie,Pettorino:2012ts,Pettorino:2013oxa,Xia:2013nua,Ade:2015rim,Miranda:2017rdk}. On the other hand, coupled dark energy models with a disformal coupling between dark matter and dark energy have been recently confronted with background cosmological data sets in Ref. \cite{vandeBruck:2016hpz}. Similar cosmological models which make use of a disformal coupling have been discussed in Refs. \cite{Zuma5,Zumalacarregui:2012us,Koivisto,Koivisto1,Sakstein:2014isa,Carsten,Sakstein:2014aca,vandeBruck:2015rma,vandeBruck:2016jgg,Brax:2016did,vandeBruck:2016cnh}. We here re--examine and update the constraints in these coupled dark energy scenarios.

The organization of this paper is as follows. In section \ref{sec:model} we introduce the generic coupled dark energy model, and in section \ref{sec:Data Sets} we summarize the observational data sets together with the method that will be employed to infer the cosmological parameter constraints. We then present the derived constraints in each coupling scenario considered in our analyses in section \ref{sec:results}. We draw our final remarks and prospective lines of research in section \ref{sec:conclusions}. 


\section{Interacting Dark Energy Model}
\label{sec:model}
In this section we briefly review the basic equations for our generic coupled dark energy (DE) model, which has been thoroughly studied in Refs. \cite{Mifsud:2017fsy,Jack}. The Einstein frame scalar--tensor theory action of our model reads
\begin{equation}\label{action}
\begin{split}
\mathcal{S} =& \int d^4 x \sqrt{-g} \left[ \frac{M_{\rm Pl}^2}{2} R - \frac{1}{2} g^{\mu\nu}\partial_\mu \phi\, \partial_\nu \phi - V(\phi) + \mathcal{L}_{SM}\right]\\
&+ \int d^4 x \sqrt{-\tilde{g}} \mathcal{\tilde{L}}_{DM}\left(\tilde g_{\mu\nu}, \psi\right),
\end{split}
\end{equation}
in which the gravitational sector has the standard Einstein--Hilbert form, and define $M_\text{Pl}^{-2}\equiv 8\pi G$ such that $M_\text{Pl}=2.4\times 10^{18}$ GeV is the reduced Planck mass. DE is described by a canonical quintessence scalar field $\phi$, with a potential $V(\phi)$. The uncoupled standard model (SM) particles are described by the Lagrangian $\mathcal{L}_{SM}$, which includes a relativistic sector $(r)$, and a baryon sector $(b)$. Particle quanta of the dark matter (DM) fields $\psi$, follow the geodesics defined by the metric
\begin{equation}\label{disformal_relation}
\tilde g_{\mu\nu} = C(\phi) g_{\mu\nu} + D(\phi)\, \partial_\mu\phi\, \partial_\nu \phi\;, 
\end{equation}
with $C(\phi),\;D(\phi)$ being the conformal and disformal coupling functions, respectively. As a consequence of the interaction between DM and DE, the conservation of the energy-momentum tensors of the scalar field and DM become
\begin{equation}\label{eq:modKG}
\Box\phi=V_{,\phi} - Q\;,\;\;\;\nabla^\mu T^{DM}_{\mu\nu}=Q\nabla_\nu\phi\;,
\end{equation}
where $V_{,\phi}\equiv dV/d\phi$, and $T^{DM}_{\mu\nu}$ is the perfect fluid energy--momentum tensor of pressureless DM. The generic coupling function is given by
\begin{equation}\label{coupling}
Q=\frac{C_{,\phi}}{2C}T_{DM}+\frac{D_{,\phi}}{2C}T_{DM}^{\mu\nu}\nabla_\mu\phi\nabla_\nu\phi-\nabla_\mu\left[\frac{D}{C}T^{\mu\nu}_{DM}\nabla_\nu\phi\right],
\end{equation}
with $T_{DM}$ being the trace of  $T_{DM}^{\mu\nu}$. In our model, SM particles are not interacting directly with the quintessence scalar field, thus their perfect fluid energy--momentum tensor satisfies $\nabla^\mu T^{SM}_{\mu\nu}=0$.

Throughout this paper, we assume a flat Universe described by the Friedmann-Robertson-Walker (FRW) line element $ds^2 = g_{\mu\nu}dx^{\mu} dx^{\nu} = a^2(\tau)\left[-d\tau^2 + \delta_{ij} dx^i dx^j\right]$, where $a(\tau)$ is the cosmological scale factor with conformal time $\tau$. In this setting, the scalar field evolves according to a modified Klein--Gordon equation
\begin{equation}\label{KG-equation}
\phi^{\prime\prime} + 2 \mathcal{H} \phi^{\prime} + a^2 V_{,\phi} = a^2 Q\;,
\end{equation}
where a prime denotes a conformal time derivative, and define the conformal Hubble parameter by $\mathcal{H}=a^\prime/a$. The DM energy density $\rho_c$, does not follow the standard redshift evolution of $a^{-3}$, but is found to satisfy an energy exchange equation
\begin{equation}\label{conservation_matter}
\rho_c^\prime + 3\mathcal{H}\rho_c = -Q\phi^{\prime}\;.
\end{equation}
In FRW, the generic coupling function simplifies to \cite{Zumalacarregui:2012us,Mifsud:2017fsy,Jack}
\begin{equation}\label{Q}
Q=\frac{2D\left(\frac{C_{,\phi}}{C}{\phi^\prime}^2 + a^2V_{,\phi} + 3\mathcal{H}\phi^\prime\right) - a^2C_{,\phi} - D_{,\phi}{\phi^\prime}^2}{2\left[a^2C + D\left(a^2\rho_c - {\phi^\prime}^2\right)\right]} \rho_c .
\end{equation}
Moreover, the radiation and baryonic energy densities satisfy the standard energy conservation equations
\begin{eqnarray}
\rho_r^\prime + 4\mathcal{H}\rho_r &=& 0\;, \\
\rho_b^\prime + 3\mathcal{H}\rho_b &=& 0\;,
\end{eqnarray}
respectively. The Friedmann equations take their usual form
\begin{eqnarray}
\mathcal{H}^2 &=& \frac{\kappa^2}{3}a^2\left(\rho_\phi + \rho_b + \rho_r + \rho_c \right)\;,\label{Friedmann} \\
\mathcal{H}^\prime &=& -\frac{\kappa^2}{6} a^2 \left(\rho_\phi + 3p_{\phi} + \rho_b + 2\rho_r + \rho_c\right)\;,
\end{eqnarray}
where $\kappa\equiv M_\text{Pl}^{-1}$, and the scalar field's energy density and pressure have the usual forms of $\rho_\phi={\phi^\prime}^2/\left(2a^2\right)+V(\phi)$ and $p_\phi=\rho_\phi-2V(\phi)$, respectively.

To be concrete, throughout this paper we choose an exponential functional form for the couplings and scalar field potential
\begin{equation}\label{coupling_choice}
C(\phi)=e^{2\alpha\kappa\phi},\;D(\phi)=D_M^4 e^{2\beta\kappa\phi},\;V(\phi)=V_0^4 e^{-\lambda\kappa\phi},
\end{equation}
where $\alpha,\,D_M,\,\beta,\,V_0,$ and $\lambda$ are constants. 

This direct interaction between DM and DE modifies both the background dynamics, as well as the evolution of perturbations \cite{Mifsud:2017fsy,Jack,Zuma2,Zuma3,Maccio:2003yk,Amendola:2003wa,TocchiniValentini:2001ty,Amendola:1999er}. For instance, these dark sector couplings modify the cosmological distances, such as the distance to the last--scattering surface, thus have a direct impact on the CMB temperature power spectrum. In addition, this interaction within the dark sector shifts the epoch of matter--radiation equality, which in turn affects the theoretical matter power spectrum.

In the rest of the paper, we will consider three main cases of the generic interacting DE model presented above, which we shall refer to as the conformal, disformal, and mixed models. For the sake of clarity, each of these cases is dealt with separately. Henceforth, in the conformal model we only consider the conformal coupling, for the disformal model we set the conformal coupling to unity and study only the disformal coupling, whereas in the mixed model we simultaneously consider the conformal and disformal couplings. The conformal coupling model has been widely studied and tight constraints were obtained from several cosmological probes \cite{Miranda:2017rdk,Ade:2015rim,Pettorino:2013oxa,Xia:2013nua,Pettorino:2012ts,Amendola:2011ie,Xia:2009zzb,Bean:2008ac,Amendola:2003eq}. A preliminary parameter constraint analysis on the couplings given in Eq. (\ref{coupling_choice}) was presented in Ref. \cite{vandeBruck:2016hpz}, which we now extend.

\section{Cosmological Data Sets \& Method}
\label{sec:Data Sets}

%
{\setlength\extrarowheight{5pt}
\setlength{\tabcolsep}{8.5pt}
\begin{table}
\begin{center}
\begin{tabular}{ l  l } 
 \hline
\hline
Parameter~  & ~Prior~ \\ 
\hline
$\Omega_b h^2$\dotfill & $[0.005,\,0.1]$ \\
$\Omega_c h^2$\dotfill & $[0.01,\,0.99]$ \\
100 $\theta_s$\dotfill & $[0.5,\,10]$ \\
$\tau_{\mathrm{reio}}$\dotfill & $[0.04,\,0.8]$ \\
$\ln(10^{10}A_s)$\dotfill & $[2.7,\,4]$ \\
$n_s$\dotfill & $[0.5,\,1.5]$ \\ 
$\lambda$\dotfill & $[0,\,1.7]$ \\ 
$\alpha$\dotfill & $[0,\,0.48]$ \\
$D_M/\,\mathrm{meV}^{-1}$\ldots & $[0,\,1.1]$ \\
$\beta$\dotfill & $[0,\,3]$ \\
\hline
\hline
\end{tabular}
\end{center}
\caption{\label{table:priors} External generic flat priors on the cosmological parameters, including the coupling dependent parameters assumed in this paper. }
\end{table}}
\begin{figure*}[t!]
\centering
  \includegraphics[width=0.98\textwidth]{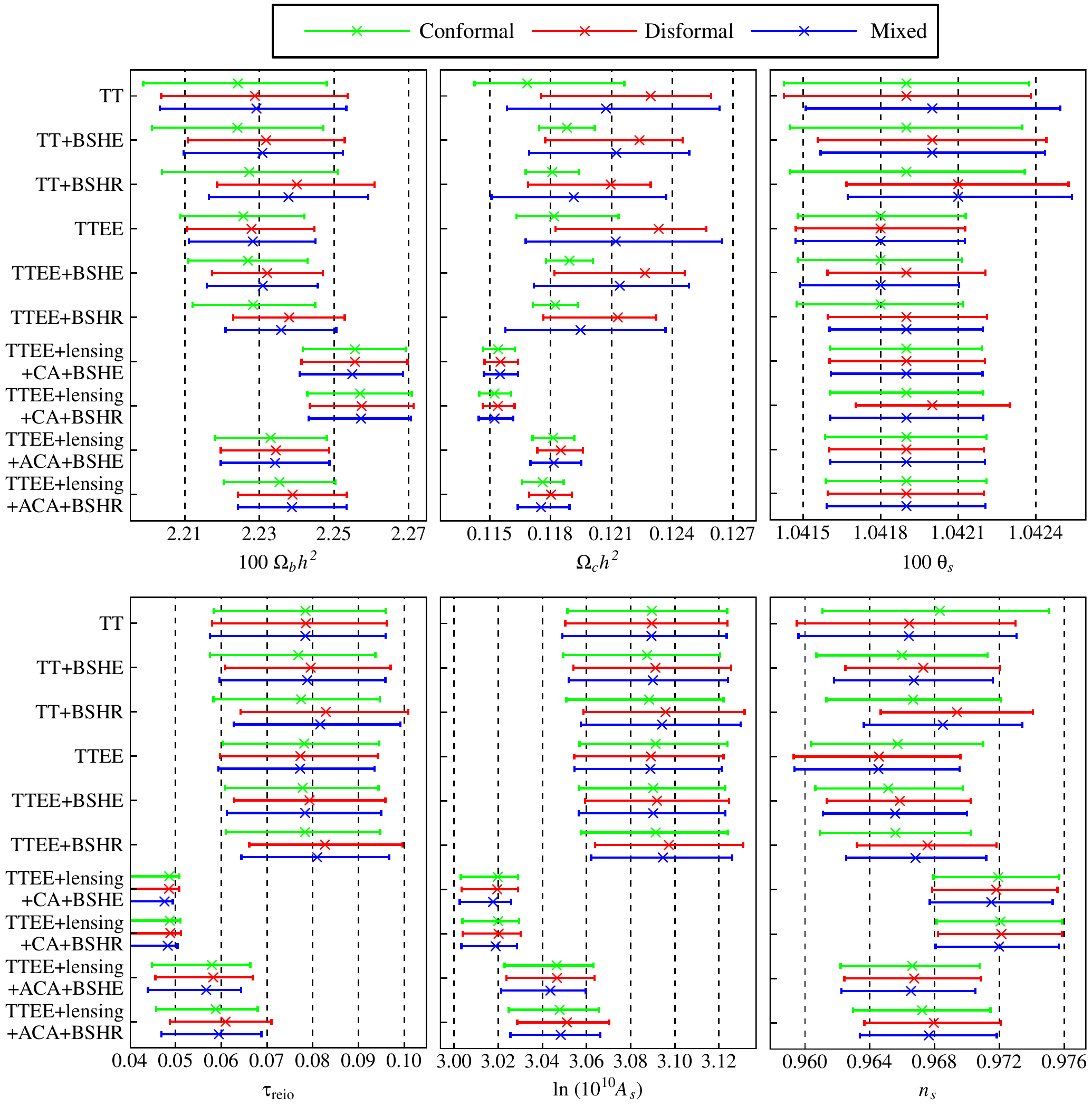}
\caption{Cosmological parameter constraints in three coupled DE models with all data set combinations considered in this paper. The coloured intervals correspond to the marginalized $1\sigma$ two--tail limits of each parameter. }  
\label{fig:dot_plot}
\end{figure*}

In the following we discuss the observational data sets that we will use in our analyses. We will be confronting the coupled DE models with probes that survey the late--time Universe, together with early--time Universe probes. 

\begin{itemize}[topsep=0pt,leftmargin=*]
  \item[] \textbf{Cosmic Microwave Background} In all data set combinations we make use of the low multipole $(2\leq\ell\leq29)$ publicly available \textit{Planck} 2015 data, which also includes the power spectra of the CMB temperature and polarization fluctuations \cite{Aghanim:2015xee}, as well as the lensing power spectrum \cite{Ade:2015zua}. For the high multipole $(l\geq30)$ range, we asses the impact of the polarization data by making use of the TT and TTTEEE likelihoods, which we denote by TT and TTEE, respectively. Occasionally we further use the \textit{Planck} lensing likelihood in the multipole range $40\leq\ell\leq400$, and we refer to this data set as lensing. 
  \item[] \textbf{Background Data} In addition, we make use of two background data set combinations which enable us to break parameter degeneracies from CMB measurements. These combinations include baryon acoustic oscillations (BAO) measurements, a supernovae Type Ia (SNIa) sample, a cosmic chronometers data set, and local measurements of the Hubble constant: 
	\begin{itemize}[topsep=0pt,leftmargin=0.5cm]
  		\item[] \textbf{Baryon Acoustic Oscillations} We consider BAO data from the SDSS Main Galaxy Sample at $z_{\mathrm{eff}}=0.15$ \cite{Ross:2014qpa}, the six degree Field Galaxy Survey at $z_{\mathrm{eff}}=0.106$ \cite{Beutler:2011hx}, and the Baryon Oscillation Spectroscopic Survey LOWZ and CMASS samples at $z_{\mathrm{eff}}=0.32$ and $z_{\mathrm{eff}}=0.57$ \cite{Cuesta:2015mqa}, respectively. 
  		\item[] \textbf{Supernovae} We make use of the SDSS--II/ SNLS3 Joint Light--curve Analysis data compilation \cite{Betoule:2014frx} of SNIa measurements.  
  		\item[] \textbf{Cosmic Chronometers} We use the compiled measurements \cite{Simon:2004tf,Stern:2009ep,Zhang:2012mp,Moresco:2012jh,Moresco:2015cya,Moresco:2016mzx} of the Hubble parameter $H(z)=a^{-1}\mathcal{H}(z)$, from the cosmic chronometers approach listed in Ref. \cite{Moresco:2016mzx}, which span the redshift range $0<z<2$.  
  		\item[] \textbf{Local Hubble Constant} In order to asses the impact of a local measurement of the Hubble constant on our coupling parameter constraints, we use the measurements as reported by Riess et al. (hereafter denoted by $H_0^{\mathrm{R}}$) \cite{Riess:2016jrr} and Efstathiou (hereafter denoted by $H_0^{\mathrm{E}}$) \cite{Efstathiou:2013via}. The choice of these measurements is motivated by the recent claims of some tension \cite{Cardona:2016ems,Bernal:2016gxb,Wu:2017fpr} within the concordance cosmological model between the CMB inferred constraint on $H_0$ and the local measurement $H_0^{\mathrm{R}}$, whereas the measurement $H_0^{\mathrm{E}}$ is found to be in very good agreement with early--Universe probes. 
	\end{itemize}
In the rest of the paper, we denote the background data set combinations $\mathrm{BAO}+\mathrm{SNIa}+H(z)+H_0^{\mathrm{E}}$ and $\mathrm{BAO}+\mathrm{SNIa}+H(z)+H_0^{\mathrm{R}}$ by BSHE and BSHR, respectively.
  \item[] \textbf{Cluster Abundance} We use cluster abundance measurements \cite{Vikhlinin:2008ym,Ade:2013lmv,Mantz:2009fw,Hajian:2013rhm,Benson:2011uta,Henry:2008cg,Rozo:2009jj,Tinker:2011pv} as a probe of the large scale structure.  This data set consists of eight measurements \cite{Bernal:2015zom} in the form of $\sigma_8(\Omega_m/\tilde{\alpha})^{\tilde{\beta}}$, where $\sigma_8\equiv\sigma_8(z=0)$ denotes the linear theory rms fluctuation in total matter in $8\,h^{-1}\,\mathrm{Mpc}$ spheres, $\Omega_m$ denotes the current total fractional abundance of matter, and the parameters $\tilde{\alpha}$ and $\tilde{\beta}$ are determined from each reported measurement. We split this data set into two measurements \cite{Vikhlinin:2008ym,Ade:2013lmv} which were found to be in tension with the concordance model (hereafter denoted by Cluster Abundance (CA)), and another subset containing the remaining six measurements \cite{Mantz:2009fw,Hajian:2013rhm,Benson:2011uta,Henry:2008cg,Rozo:2009jj,Tinker:2011pv} (hereafter denoted by Alternative Cluster Abundance (ACA)). Although in the analyses that follow we do not report the cosmological parameter constraints obtained from a joint analysis of the CA and ACA data sets, we have checked that in our coupled DE models, the derived constraints from a joint analysis are in an excellent agreement with the results from the CA data set analysis. This is because the two measurements found in the CA data set have the smallest error--bars and thus they dominate in a joint analysis. Moreover, we should mention that these cluster abundance measurements should be taken with a pinch of salt, due to their dependence on the concordance model under which these measurements were inferred. However, our goal is to check if the individual data sets can be brought in good agreement with each other with the inclusion of the DE interactions. 
\end{itemize}

We employ a Bayesian approach to infer the parameter posterior distributions together with their confidence limits. This is implemented by the Markov Chain Monte Carlo (MCMC) technique via a customized version of \texttt{Monte Python} \cite{Audren:2012wb} which is interfaced with a modified version of the cosmological Boltzmann code \texttt{CLASS} \cite{Blas:2011rf}. Apart from the implementation of our generic coupled DE model equations, we also included a shooting algorithm in \texttt{CLASS} in order to find the scalar field potential energy scale $V_0$. The equations governing the evolution of perturbations \cite{Mifsud:2017fsy} in our generic coupled DE model were implemented in both the Newtonian and synchronous gauge, and verified that we get identical results in the two gauges. For all the models considered in sections \ref{sec:conformal_results}--\ref{sec:mixed_results}, we also made use of the MCMC analysis package \texttt{GetDist} \cite{Lewis:2002ah}, and checked that the results are in an excellent agreement with those obtained from \texttt{Monte Python}.
\\\hspace*{1em}We consider flat priors for the generic cosmological parameters that are allowed to vary in our MCMC analyses. The full range of each flat prior is listed in Table \ref{table:priors}. The general baseline set of parameters consists of $\Theta=\{\Omega_b h^2,\,\Omega_c h^2,\,100\,\theta_s,\,\tau_{\mathrm{reio}},\,\ln(10^{10}A_s),\,n_s,\,\lambda,\,\alpha,\,D_M,\,\beta\}$. Here, $h$ is defined in terms of the Hubble constant via $H_0=100\,h\,\mathrm{km}\,\mathrm{s}^{-1}\mathrm{Mpc}^{-1}$, $\Omega_b h^2$ represents the effective fractional abundance of uncoupled baryons, $\Omega_c h^2$ is the pressureless coupled cold dark matter effective energy density, $100\,\theta_s$ is the angular scale of the sound horizon at last scattering defined by the ratio of the sound horizon at decoupling to the angular diameter distance to the last scattering surface, $\tau_{\mathrm{reio}}$ is the reionization optical depth parameter, $\ln(10^{10}A_s)$ is the log power of the scalar amplitude of the primordial power spectrum together with its scalar spectral index $n_s$, $\lambda$ is the slope of the scalar field exponential potential, $\alpha$ is the conformal coupling parameter, and $D_M$ is the energy scale of the disformal coupling together with the disformal exponent $\beta$. The pivot scale in our analyses was set to $k_0=0.05\,\textrm{Mpc}^{-1}$, and we assume purely adiabatic scalar perturbations at very early times without the running of the scalar spectral index. Moreover, we fix the neutrino effective number to its standard value of $N_\mathrm{eff}=3.046$ \cite{Mangano:2001iu}, as well as the photon temperature today to $T_0=2.7255\,\mathrm{K}$ \cite{Fixsen:2009ug}. As mentioned earlier, we assume spatial flatness.

In the top block of Tables \ref{table:conf_Tab1}--\ref{table:mixed_Tab2}, we present the constraints on the parameters with flat priors that are varied in the MCMC analyses of the respective coupled DE model. In our analyses, we also consider marginalized constraints on various derived parameters which we present in the lower block of Tables \ref{table:conf_Tab1}--\ref{table:mixed_Tab2}. The derived parameters include today's value of the Hubble parameter $H_0$ in $\mathrm{km}\,\mathrm{s}^{-1}\mathrm{Mpc}^{-1}$, $\Omega_m$, $\sigma_8$, the reionization redshift $z_\mathrm{reio}$, and the dimensionless age of the Universe $H_0t_0$, with $t_0$ being the current age of the Universe.    

\section{Results}
\label{sec:results}
In this section we discuss the inferred cosmological parameter constraints following the procedure described in section \ref{sec:Data Sets}. We first consider a coupled DE model with an exponential conformal coupling only, which we discuss in section \ref{sec:conformal_results}, and then we present the obtained constraints in the constant as well as in the exponential disformally coupled DE models in section \ref{sec:disformal_results}. Finally, in section \ref{sec:mixed_results} we discuss the derived constraints for the mixed coupled model which simultaneously makes use of both the exponential conformal and constant disformal couplings between the dark sector constituents. In the mixed model, we further consider a particular case in which we fix the constant disformal coupling parameter $D_M$, in order to assess the impact on the conformal coupling parameter constraint.    

\begin{figure}
\centering
  \includegraphics[width=0.98\columnwidth]{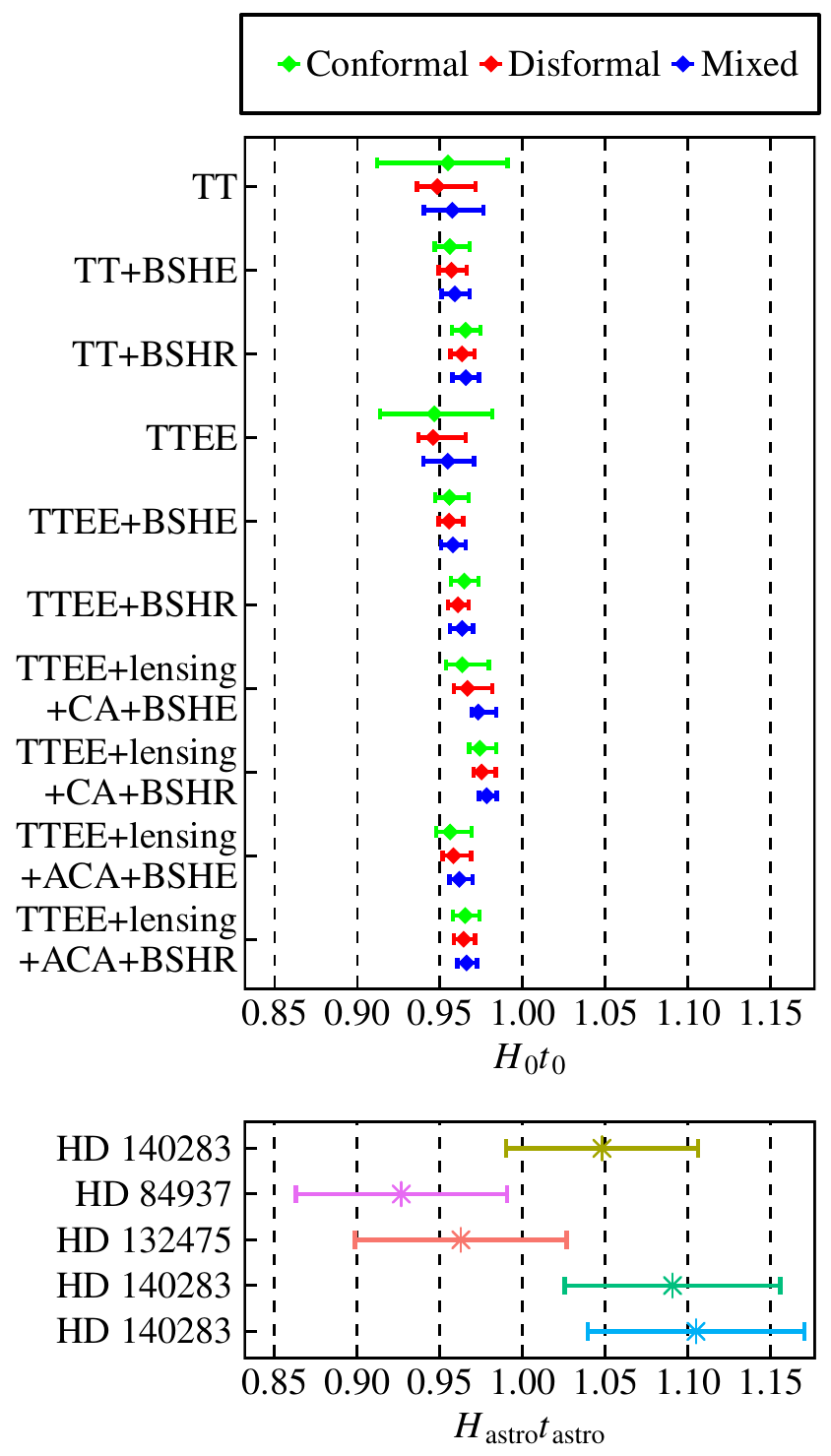}
\caption{In the upper panel we compare the marginalized constraints on the dimensionless age of the Universe in the conformal, disformal, and mixed coupled DE models using all data set combinations considered in this paper. The lower panel depicts the constraints on $H_\mathrm{astro}t_\mathrm{astro}$ from astrophysical objects \cite{refId0,0004-637X-792-2-110,2041-8205-765-1-L12} with their names specified on the vertical axis. In the upper panel the coloured intervals correspond to the inferred $1\sigma$ two--tail limits on the dimensionless age of the Universe, whereas in the lower panel these intervals show the estimated $1\sigma$ constraints.}  
\label{fig:H0t0_plot}
\end{figure}
{\setlength\extrarowheight{5pt}
\begin{table*}
\begin{center}
\begin{tabular}{ l c  c  c  c  c  c} 
 \hline
\hline
Parameter~  &  ~TT~ & ~$\mathrm{TT+BSHE}$~ & ~$\mathrm{TT+BSHR}$~ & ~TTEE~ & ~$\mathrm{TTEE+BSHE}$~ & ~$\mathrm{TTEE+BSHR}$~  \\
\hline
100 $\Omega_b h^2$\dotfill & $2.2242^{+0.023823}_{-0.025299}$ & $2.2241^{+0.023010}_{-0.022874}$ & $2.2273^{+0.023649}_{-0.023425}$ & $2.2257^{+0.016353}_{-0.016872}$ & $2.2269^{+0.015977}_{-0.015975}$ & $2.2283^{+0.016622}_{-0.016223}$ \\
$\Omega_c h^2$\dotfill & $0.11685^{+0.0047878}_{-0.0026041}$ & $0.11880^{+0.0013776}_{-0.0013724}$ & $0.11808^{+0.0013206}_{-0.0013136}$ & $0.11817^{+0.0031795}_{-0.0018489}$ & $0.11893^{+0.0011569}_{-0.0011555}$ & $0.11823^{+0.0011171}_{-0.0011011}$ \\
100 $\theta_s$\dotfill & $1.0419^{+0.00047486}_{-0.00047338}$ & $1.0419^{+0.00044728}_{-0.00045081}$ & $1.0419^{+0.00045730}_{-0.00044997}$ & $1.0418^{+0.00032875}_{-0.00031963}$ & $1.0418^{+0.00031572}_{-0.00031956}$ & $1.0418^{+0.00031864}_{-0.00032426}$ \\
$\tau_{\mathrm{reio}}$\dotfill & $0.078411^{+0.017508}_{-0.020092}$ & $0.076849^{+0.016848}_{-0.019297}$ & $0.077484^{+0.017183}_{-0.019231}$ & $0.078153^{+0.016437}_{-0.017801}$ & $0.077784^{+0.016594}_{-0.017016}$ & $0.078322^{+0.016436}_{-0.017309}$ \\
$\ln(10^{10}A_s)$\ldots & $3.0896^{+0.034168}_{-0.038354}$ & $3.0874^{+0.033060}_{-0.037978}$ & $3.0884^{+0.033680}_{-0.037607}$ & $3.0913^{+0.032600}_{-0.034479}$ & $3.0902^{+0.032568}_{-0.033662}$ & $3.0913^{+0.032653}_{-0.033779}$ \\
$n_s$\dotfill & $0.96832^{+0.0067539}_{-0.0072504}$ & $0.96598^{+0.0052916}_{-0.0052853}$ & $0.96668^{+0.0054260}_{-0.0053638}$ & $0.96571^{+0.0052979}_{-0.0053330}$ & $0.96514^{+0.0045934}_{-0.0045062}$ & $0.96558^{+0.0046446}_{-0.0046646}$ \\ 
$\lambda$\dotfill & $<1.2170(1.6013)$ & $<0.6686(1.0133)$ & $<0.4528(0.8046)$ & $<1.1718(1.5981)$ & $<0.6228(0.9927)$ & $<0.4481(0.7957)$ \\ 
$\alpha$\dotfill & $<0.0582(0.1037)$ & $<0.0360(0.0543)$ & $0.032032^{+0.019815}_{-0.017833}$ & $<0.0496(0.0881)$ & $<0.0394(0.0519)$ & $0.032964^{+0.019626}_{-0.014047}$ \\
\hline
$H_0$\dotfill & $68.373^{+2.8145}_{-3.9906}$ & $68.031^{+0.91665}_{-0.80492}$ & $68.848^{+0.76372}_{-0.78577}$ & $67.553^{+2.7463}_{-2.9482}$ & $68.006^{+0.88745}_{-0.78281}$ & $68.786^{+0.73978}_{-0.77769}$ \\
$\Omega_m$\dotfill & $0.30053^{+0.040750}_{-0.032515}$ & $0.30491^{+0.0090961}_{-0.0098531}$ & $0.29624^{+0.0084333}_{-0.0086687}$ & $0.30976^{+0.030719}_{-0.030543}$ & $0.30546^{+0.0088394}_{-0.0093426}$ & $0.29711^{+0.0084298}_{-0.0083116}$ \\
$\sigma_8$\dotfill & $0.84733^{+0.030431}_{-0.043860}$ & $0.84332^{+0.018697}_{-0.020182}$ & $0.85064^{+0.018667}_{-0.022011}$ & $0.84285^{+0.028084}_{-0.035870}$ & $0.84441^{+0.017058}_{-0.017744}$ & $0.85223^{+0.017132}_{-0.018724}$ \\
$z_{\mathrm{reio}}$\dotfill & $9.8999^{+1.6689}_{-1.6576}$ & $9.8048^{+1.6436}_{-1.6258}$ & $9.8424^{+1.6665}_{-1.6003}$ & $9.9114^{+1.5890}_{-1.4639}$ & $9.8945^{+1.5883}_{-1.4342}$ & $9.9287^{+1.5790}_{-1.4473}$ \\
$H_0 t_0$\dotfill & $0.95482^{+0.036012}_{-0.042848}$ & $0.95592^{+0.011935}_{-0.009077}$ & $0.96538^{+0.0090549}_{-0.0082628}$ & $0.94652^{+0.035001}_{-0.032737}$ & $0.95571^{+0.011533}_{-0.008620}$ & $0.96467^{+0.0086895}_{-0.0081049}$ \\
\hline
\hline
\end{tabular}
\end{center}
\caption{\label{table:conf_Tab1} For each model parameter we report the mean values and $1\sigma$ errors in the conformally coupled DE scenario. The Hubble constant is given in units of $\mathrm{km}\,\mathrm{s}^{-1}\,\mathrm{Mpc}^{-1}$. When necessary, for the model parameters $\lambda$ and $\alpha$, we also write in brackets the $2\sigma$ upper limits.}
\end{table*}}
\begin{figure*}
\centering
  \includegraphics[width=0.98\textwidth]{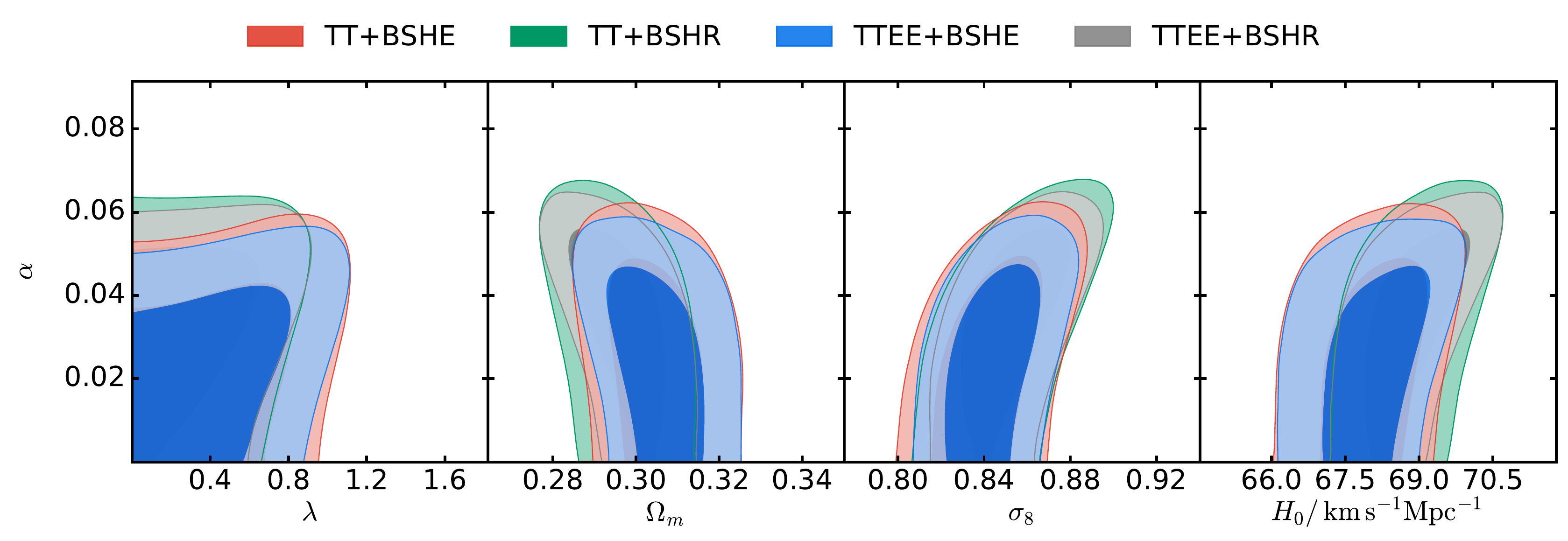}
\caption{Marginalized two--dimensional likelihood constraints for conformally coupled DE with different data set combinations. We show the degeneracy of the conformal coupling parameter $\alpha$ with $\lambda,\,\Omega_m,\,\sigma_8,$ and $H_0$.}  
\label{fig:conf_2D_Tab1}
\end{figure*}

In Fig. \ref{fig:dot_plot} we illustrate the obtained cosmological parameter constraints on the usual six varied parameters of the concordance model from the MCMC likelihood analyses in the conformal, disformal, and mixed coupled DE models with all data set combinations considered throughout this paper. For the disformal model we only show the inferred constraints from the constant disformally coupled model, since the $1\sigma$ limits do not change appreciably in the exponential disformally coupled case presented in Table \ref{table:disf_exp_Tab}. Similarly, for the mixed coupled model we do not show the constraints from the mixed model with fixed $D_M$, considered in the last column of Table \ref{table:mixed_Tab2}. 

A generic feature of our coupled DE models is that when the cluster abundance data sets are included, $\tau_{\mathrm{reio}}$ and $A_s$ are shifted to lower values in comparison with their inferred mean values from the other data set combinations. This shift is predominantly observed when using the CA data set rather than the ACA measurements. The reason behind this is that the measurements contained in the CA data set prefer lower values of $\sigma_8$ in all coupled DE models with respect to the other data sets, including those in the ACA data set. Moreover, the major impacts of this shift in the range of $\sigma_8$ are found to be on $\tau_{\mathrm{reio}}$ and $A_s$, which follow from the degeneracies between $\tau_\mathrm{{reio}}$ and $\sigma_8$, and between $A_s$ and $\sigma_8$. As a consequence of these degeneracies, a discrepancy between the $1\sigma$ limits of $\tau_{\mathrm{reio}}$ and $A_s$ arises between the data set combinations which make use of the CA measurements with the other data set combinations which do not include the cluster abundance measurements. This is clearly shown in Fig. \ref{fig:dot_plot}. Nonetheless, the inferred values of $\tau_{\mathrm{reio}}$ from all data set combinations, including those combinations which use the cluster abundance data sets, are still in agreement with constraints from other reionization probes \cite{Robertson:2015uda,Becker:2015lua,McQuinn:2015icp}. Clearly, improved accuracy on the reionization optical depth parameter will be useful to break the degeneracies with other cosmological parameters \cite{Aghanim:2016yuo,Adam:2016hgk}. Furthermore, we should also mention that there is a partial inverse correlation between $\sigma_8$ and $n_s$. Thus, the $1\sigma$ limits on $n_s$ shift to slightly larger values for the data set combinations which include the CA measurements with respect to the other data sets.

In the upper panel of Fig. \ref{fig:H0t0_plot}, we show the inferred constraints on the dimensionless age of the Universe in the models presented in Fig. \ref{fig:dot_plot}, and in the lower panel we show the $1\sigma$ intervals from astrophysical objects. For the calculation of the dimensionless age of the Universe $H_\mathrm{astro}t_\mathrm{astro}$, we use the estimation of the astrophysical age of the Universe based on some of the best known oldest stars \cite{refId0,0004-637X-792-2-110,2041-8205-765-1-L12}, and assume the value of the astrophysical Hubble constant to coincide with $H_0^{\mathrm{R}}$. We should emphasize that the $H_\mathrm{astro}t_\mathrm{astro}$ constraints in Fig. \ref{fig:H0t0_plot} are solely used for comparative purposes and not in our cosmological parameter constraints analyses. The comparison of the constraints presented in the upper and lower panels of Fig. \ref{fig:H0t0_plot} can be interpreted as a convergence between the theory of General Relativity which governs the cosmological evolution of the Universe, and the laws of quantum mechanics which determine the nuclear reactions taking place in stars. Following our MCMC analyses, the present time coincidence of $H_0t_0=1$, which has been recently dubbed as the synchronicity problem \cite{Avelino:2016qjs}, is not completely fulfilled in our coupled DE models as $H_0t_0$ is not found to be exactly unity. Nonetheless, it still remains to be seen if this makes the synchronicity problem even worse \cite{Melia:2007sd,Melia:2011fj,vanOirschot:2010mc}. 

\subsection{Conformal model constraints}
\label{sec:conformal_results}

\begin{figure}
\centering
  \includegraphics[width=0.98\columnwidth]{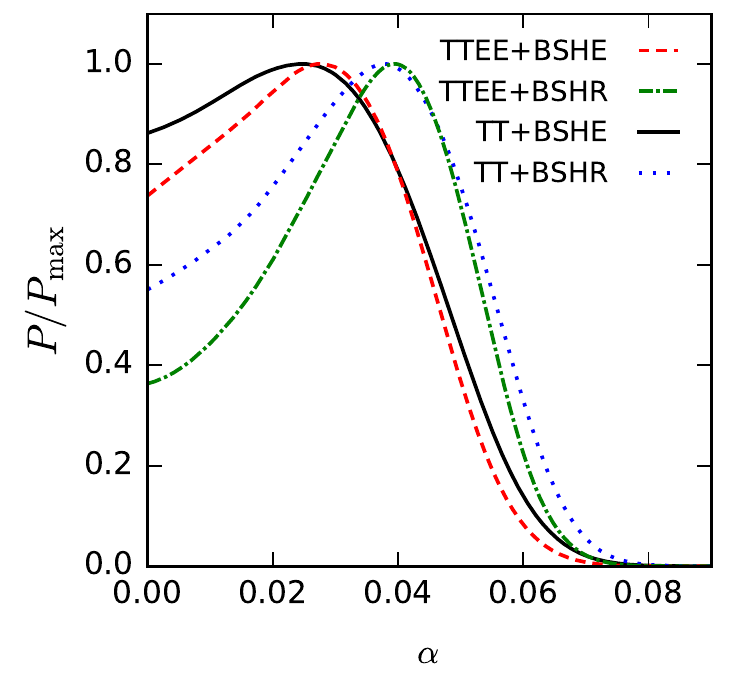}
\caption{Marginalized one--dimensional posterior distributions for the conformal coupling parameter $\alpha$, with the different data set combinations indicated in the figure. The respective parameter constraints are tabulated in Table \ref{table:conf_Tab1}.}  
\label{fig:conf_1D_Tab1}
\end{figure}

\begin{figure*}
\centering
  \includegraphics[width=0.98\textwidth]{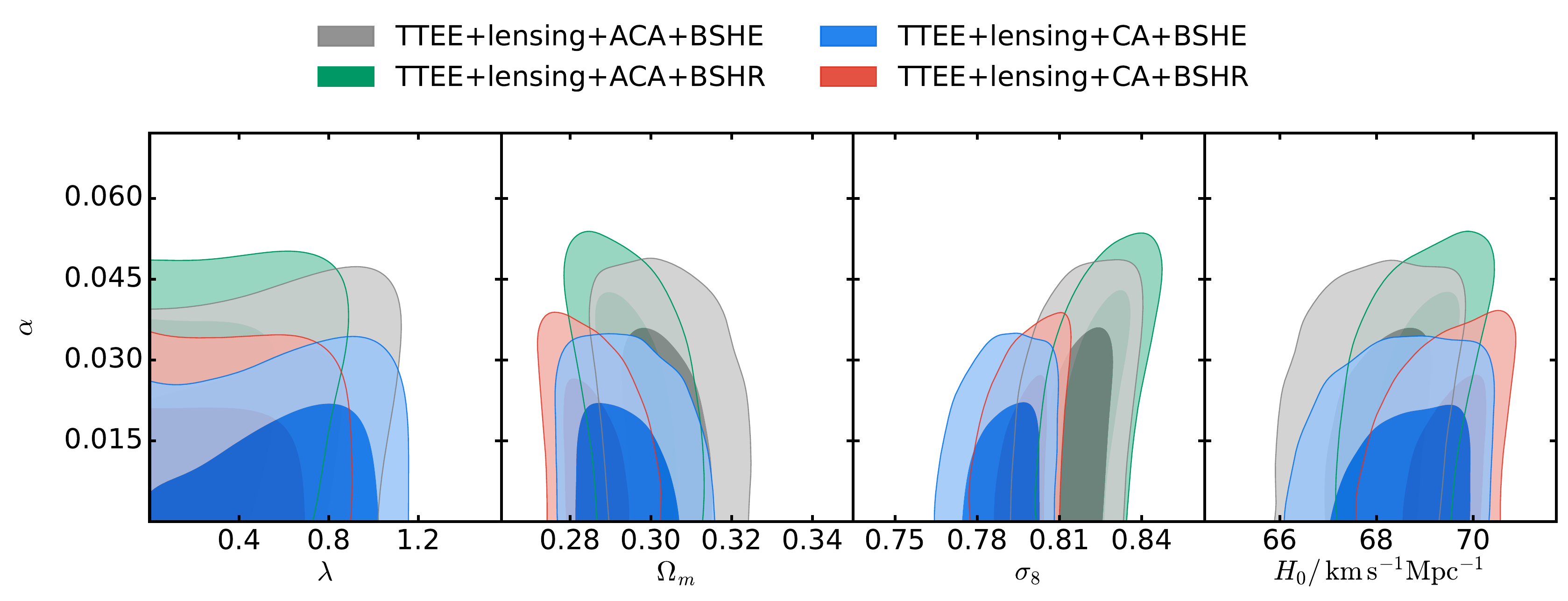}
\caption{Marginalized two--dimensional likelihood constraints on the parameters $\alpha,\,\lambda,\,\Omega_m,\,\sigma_8,$ and $H_0$ in the conformal model. The respective parameter constraints are tabulated in Table \ref{table:conf_Tab2}.}  
\label{fig:conf_2D_Tab2}
\end{figure*}

\begin{figure}
\centering
  \includegraphics[width=0.98\columnwidth]{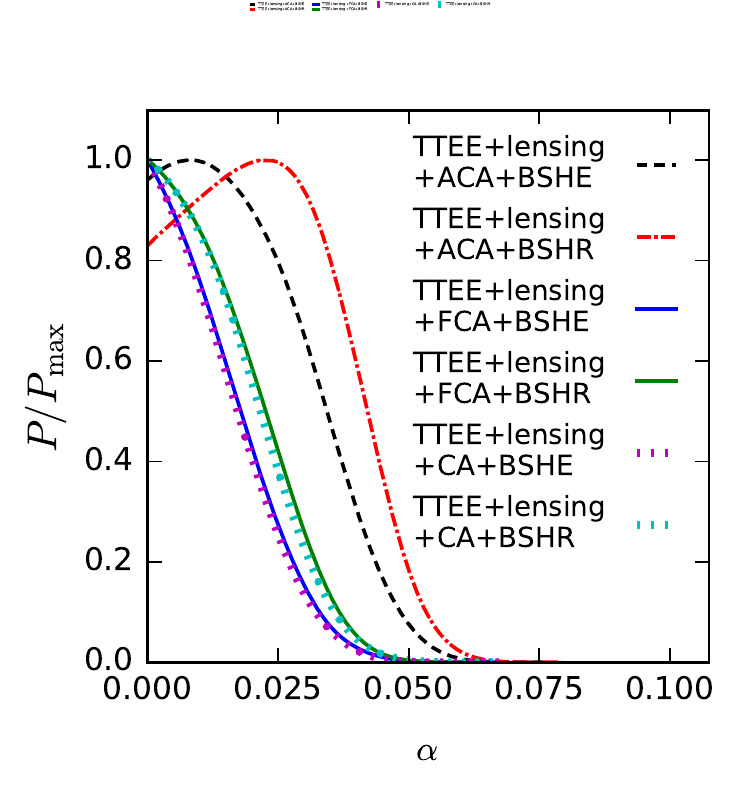}
\caption{Marginalized one--dimensional posterior distributions for the conformal coupling parameter $\alpha$, with the different data set combinations considered in Table \ref{table:conf_Tab2}, together with two other combinations making use of both the ACA and CA measurements (denoted by FCA).}  
\label{fig:conf_1D_Tab2}
\end{figure}

In this section we discuss the inferred constraints in the exponential conformally coupled model, with the coupling parameter $\alpha$ as defined in Eq. (\ref{coupling_choice}). In this model we neglect the disformal coupling by fixing $D_M$ to zero. In Tables \ref{table:conf_Tab1} and \ref{table:conf_Tab2} we tabulate the parameter constraints from several data set combinations. The marginalized two--dimensional likelihood constraints and the one--dimensional posterior distributions for the coupling parameter $\alpha$ of Table \ref{table:conf_Tab1} are shown in Fig. \ref{fig:conf_2D_Tab1} and Fig. \ref{fig:conf_1D_Tab1}, respectively. Similarly, the marginalized two--dimensional likelihood constraints and the one--dimensional posterior distributions for the coupling parameter $\alpha$ of Table \ref{table:conf_Tab2} are shown in Fig. \ref{fig:conf_2D_Tab2} and Fig. \ref{fig:conf_1D_Tab2}, respectively. As clearly illustrated in Fig. \ref{fig:dot_plot}, marginally tighter constraints on the cosmological parameters are obtained with the TTEE CMB likelihood in comparison with the TT likelihood. Consequently, the $95\%$ confidence level (C.L.) upper bound on the conformal coupling parameter decreases from $\alpha<0.1037$ with the TT likelihood, to $\alpha<0.0881$ when using the TTEE likelihood.

Since the CMB anisotropies mainly probe the high--redshift Universe, we further add some information about the low--redshift Universe from the background data sets BSHE and BSHR, which will also help to break the degeneracy between the parameters. From Table \ref{table:conf_Tab1} and Fig. \ref{fig:dot_plot}, it follows that the background data sets improve the constraints on the cosmological parameters, particularly on the current matter abundance fraction $\Omega_m$. From the second panel of Fig. \ref{fig:conf_2D_Tab1}, it is evident that there is a partial inverse correlation between $\Omega_m$ and $\alpha$, thus a lower upper bound on $\alpha$ results into a slightly higher mean value of $\Omega_m$. This clearly follows from the transfer of energy between DM and DE which is governed by the conservation equations (\ref{KG-equation}) and (\ref{conservation_matter}).

{\setlength\extrarowheight{5pt}
\begin{table*}
\begin{center}
\begin{tabular}{ l  c  c  c  c } 
 \hline
\hline
Parameter~  & ~\begin{tabular}[t]{@{}c@{}}$\mathrm{TTEE+lensing}$ \\ $\mathrm{+\,CA+BSHE}$\end{tabular}~ 
			& ~\begin{tabular}[t]{@{}c@{}}$\mathrm{TTEE+lensing}$ \\ $\mathrm{+\,CA+BSHR}$\end{tabular}~ 
			& ~\begin{tabular}[t]{@{}c@{}}$\mathrm{TTEE+lensing}$ \\ $\mathrm{+\,ACA+BSHE}$\end{tabular}~
			& ~\begin{tabular}[t]{@{}c@{}}$\mathrm{TTEE+lensing}$ \\ $\mathrm{+\,ACA+BSHR}$\end{tabular}~ \\ 
\hline
100 $\Omega_b h^2$\dotfill & $2.2556^{+0.013677}_{-0.013975}$ & $2.2570^{+0.013806}_{-0.014149}$ & $2.2330^{+0.015035}_{-0.014913}$ & $2.2354^{+0.014958}_{-0.014915}$ \\
$\Omega_c h^2$\dotfill & $0.11541^{+0.00083356}_{-0.00073576}$ & $0.11523^{+0.00081890}_{-0.00075019}$ & $0.11812^{+0.0010400}_{-0.0010158}$ & $0.11761^{+0.0010363}_{-0.0010082}$ \\
100 $\theta_s$\dotfill & $1.0419^{+0.00029123}_{-0.00029659}$ & $1.0419^{+0.00029641}_{-0.00029487}$ & $1.0419^{+0.00030897}_{-0.00031404}$ & $1.0419^{+0.00030894}_{-0.00031132}$ \\
$\tau_{\mathrm{reio}}$\dotfill & $0.048632^{+0.0022281}_{-0.0086316}$ & $0.048728^{+0.0023093}_{-0.0087274}$ & $0.057948^{+0.008405}_{-0.013138}$ & $0.058798^{+0.009195}_{-0.013012}$ \\
$\ln(10^{10}A_s)$\ldots & $3.0197^{+0.009227}_{-0.016733}$ & $3.0199^{+0.009476}_{-0.016226}$ & $3.0464^{+0.016778}_{-0.023569}$ & $3.0478^{+0.017737}_{-0.023093}$ \\
$n_s$\dotfill & $0.97192^{+0.0037326}_{-0.0039877}$ & $0.97206^{+0.0038073}_{-0.0039352}$ & $0.96662^{+0.0041491}_{-0.0044300}$ & $0.96723^{+0.0042144}_{-0.0042766}$ \\ 
$\lambda$\dotfill & $0.61752^{+0.37467}_{-0.24731}$ & $<0.5106(0.8274)$ & $<0.8175(1.0290)$ & $<0.4550(0.7961)$ \\ 
$\alpha$\dotfill & $<0.0153(0.0301)$ & $<0.0174(0.0325)$ & $<0.0247(0.0423)$ & $<0.0331(0.0467)$ \\
\hline
$H_0$\dotfill & $68.623^{+1.1966}_{-0.8159}$ & $69.460^{+0.77325}_{-0.56054}$ & $67.998^{+0.98167}_{-0.74259}$ & $68.744^{+0.69420}_{-0.68187}$ \\
$\Omega_m$\dotfill & $0.29316^{+0.007559}_{-0.010426}$ & $0.28570^{+0.0058052}_{-0.0072021}$ & $0.30392^{+0.0081011}_{-0.0096183}$ & $0.29629^{+0.0076271}_{-0.0075417}$ \\
$\sigma_8$\dotfill & $0.79003^{+0.011312}_{-0.008929}$ & $0.79698^{+0.0080525}_{-0.0070617}$ & $0.81687^{+0.010712}_{-0.010269}$ & $0.82300^{+0.009523}_{-0.010580}$ \\
$z_{\mathrm{reio}}$\dotfill & $6.9761^{+0.24398}_{-0.96532}$ & $6.9806^{+0.24979}_{-0.96634}$ & $7.9986^{+0.9093}_{-1.2186}$ & $8.0700^{+0.9729}_{-1.2031}$ \\
$H_0 t_0$\dotfill & $0.96341^{+0.016001}_{-0.009905}$ & $0.97410^{+0.0099529}_{-0.0065924}$ & $0.95609^{+0.013021}_{-0.008549}$ & $0.96517^{+0.0087182}_{-0.0072581}$ \\
\hline
\hline
\end{tabular}
\end{center}
\caption{\label{table:conf_Tab2} For each model parameter we report the mean values and $1\sigma$ errors in the conformally coupled DE scenario. The Hubble constant is given in units of $\mathrm{km}\,\mathrm{s}^{-1}\,\mathrm{Mpc}^{-1}$. When necessary, for the model parameters $\lambda$ and $\alpha$, we also write in brackets the $2\sigma$ upper limits. }
\end{table*}}

Moreover, the coupling parameter $\alpha$ is found to be marginally correlated with the Hubble constant, as depicted in the fourth panel of Fig. \ref{fig:conf_2D_Tab1}. Indeed, higher upper bounds on $\alpha$ are inferred when using the $H_0^{\mathrm{R}}$ local value of the Hubble constant in comparison with the analyses making use of $H_0^{\mathrm{E}}$. In fact, for the $\mathrm{TT+BSHR}$ and $\mathrm{TTEE+BSHR}$ combinations, we quote the peak locations in the one--dimensional posterior distributions of $\alpha$, as depicted in Fig. \ref{fig:conf_1D_Tab1}. The peak in the posterior distribution with $\mathrm{TT+BSHR}$ is found to be at $\alpha=0.032032^{+0.019815}_{-0.017833}$, whereas with $\mathrm{TTEE+BSHR}$ the peak is at $\alpha=0.032964^{+0.019626}_{-0.014047}$. Thus, a higher value of $H_0$ together with the CMB polarization likelihood enhance the preference of a non--zero $\alpha$, although in the two mentioned cases the conformal coupling parameter is still found to be consistent with zero at $\sim2\sigma$.  This complements the discussion of this model with an inverse power--law potential in Ref. \cite{Ade:2015rim}. Similar indications of a non--null coupling, although with a different coupling function, have also been reported in Ref. \cite{DiValentino:2017iww}. Also, phantom dark energy was found to be preferred when relatively high external local values of $H_0$ are adopted \cite{Shafer:2013pxa,Bonvin:2016crt}. The TT/ $\mathrm{TTEE+BSHE}$ data set combinations do not give rise to a significant peak in the marginalized posterior distribution of $\alpha$, although a tighter constraint on the conformal coupling parameter is obtained with the TTEE likelihood in comparison with the constraint from the TT likelihood combination. 

The inferred upper bound constraints on the slope of the exponential scalar field potential $\lambda$, are significantly improved when we include the background data sets BSHE and BSHR along with the CMB likelihoods. This is mainly due to the fact that the derived constraints on $\Omega_m$ are tighter with the background data sets, leading to a considerable improvement in the upper bounds of $\lambda$, which is correlated with $\Omega_m$. The background data sets lower the $95\%\;\mathrm{C.L.}$ upper bounds on $\lambda$, from $\lambda<1.5981$ with TTEE, to $\lambda<0.9927$ with $\mathrm{TTEE+BSHE}$, and particularly to $\lambda<0.7957$ with $\mathrm{TTEE+BSHR}$, all consistent with Refs. \cite{Bean:2008ac,Xia:2009zzb,Miranda:2017rdk}. We show the correlation between $\alpha$ and $\lambda$ in the first panel of Fig. \ref{fig:conf_2D_Tab1}.  

\begin{figure}
\centering
  \includegraphics[width=0.98\columnwidth]{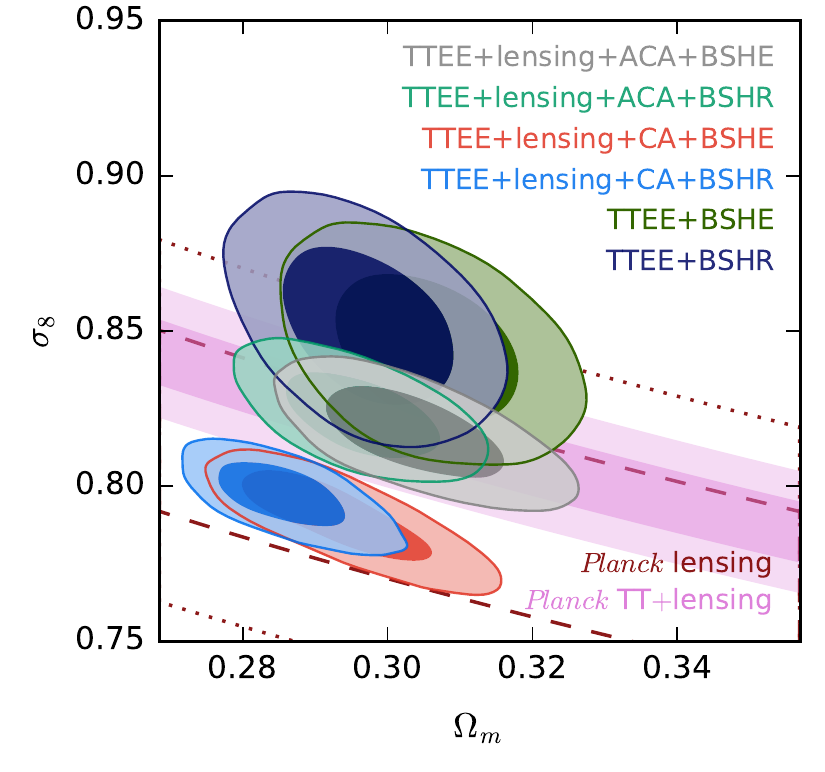}
\caption{Marginalized constraints on parameters of the conformal model using the data sets indicated in the figure. The shaded band depicts the \textit{Planck} $\mathrm{TT+lensing}$ constraint, whereas the region enclosed by the dashed $(1\sigma)$ and dotted $(2\sigma)$ lines shows the constraint from \textit{Planck} lensing alone \cite{Ade:2015zua}.}  
\label{fig:s8_Om_bands}
\end{figure}

Conformally coupled DE models are known to be characterised by higher values of $\sigma_8$ in comparison with the concordance and uncoupled quintessence models \cite{Mifsud:2017fsy,Pourtsidou:2013nha,Copeland:2006wr} as a result of an enhancement in the growth of perturbations. The correlation between the coupling parameter $\alpha$ and $\sigma_8$ is shown in the third panel of Fig. \ref{fig:conf_2D_Tab1}. In order to probe the growth of perturbations, we now consider the cluster abundance data sets, as well as the CMB gravitational lensing likelihood. In Table \ref{table:conf_Tab2} we further include the lensing, CA, and ACA data sets in our analyses, and we find that the conformal coupling parameter upper bounds are lowered in comparison with the inferred upper bounds from the data sets considered in Table \ref{table:conf_Tab1}. The two--dimensional marginalized constraints on $\alpha$ with the parameters $\lambda,\,\Omega_m,\,\sigma_8,$ and $H_0$ are shown in Fig. \ref{fig:conf_2D_Tab2}. From the marginalized posterior distributions of the conformal coupling parameter, shown in Fig. \ref{fig:conf_1D_Tab2}, we find that the observed peaks in Fig. \ref{fig:conf_1D_Tab1} are now insignificant when we include the cluster abundance and lensing data sets. In Fig. \ref{fig:conf_1D_Tab2}, we also show the marginalized posterior distributions of $\alpha$, inferred from the analyses which include the Full Cluster Abundance (FCA) data set consisting of the CA and ACA measurements altogether. As already mentioned in section \ref{sec:Data Sets}, the derived constraints on $\alpha$ from the FCA data set coincide with the obtained constraints from the CA data set, henceforth we do not report the parameter constraints from the MCMC analyses which make use of the FCA data set.

\begin{figure*}
\centering
  \includegraphics[width=0.98\columnwidth]{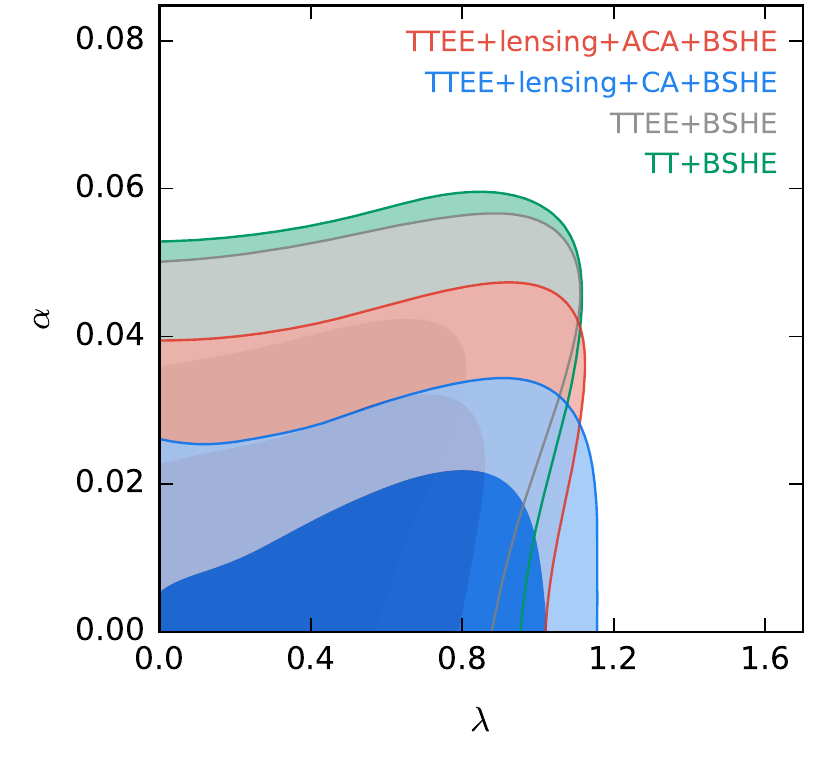}
  \includegraphics[width=0.984\columnwidth]{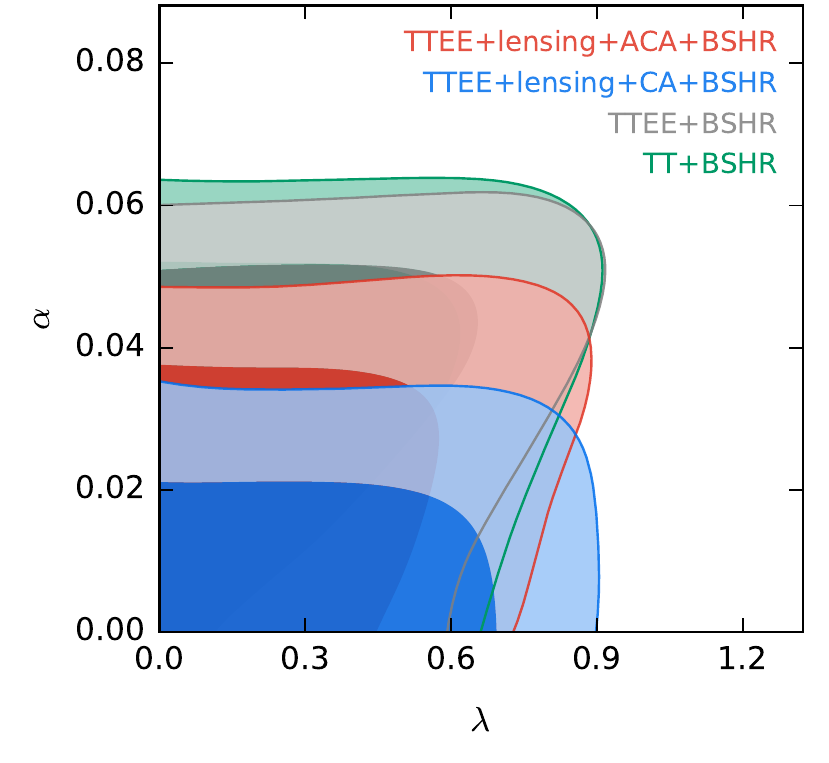}
\caption{A comparison of the marginalized two--dimensional constraints on the conformal coupling parameter $\alpha$, and the slope of the exponential potential $\lambda$, using the local values of the Hubble constant $H_0^{\mathrm{E}}$ (left) and $H_0^{\mathrm{R}}$ (right).}  
\label{fig:alpha_lambda_ER}
\end{figure*}

The tightest $95\%\;\mathrm{C.L.}$ upper bounds on $\alpha$ are derived from the CA data set combinations, since the measurements in this cluster abundance data set favour relatively low values of $\sigma_8$. In fact, these CA measurements are in tension with the inferred concordance model $\sigma_8$ constraints \cite{Ade:2015xua,Ade:2013lmv,Bernal:2015zom}. In Fig. \ref{fig:s8_Om_bands}, we show the two--dimensional marginalized constraints on $\sigma_8$ and $\Omega_m$ from two data set combinations which do not include the cluster abundance and lensing data sets, together with the data set combinations which probe the growth of perturbations. For comparative purposes only, we also include the concordance model constraints inferred from the CMB lensing only likelihood (depicted by dashed and dotted lines), and from the CMB $\mathrm{TT+lensing}$ likelihoods (depicted by the shaded bands) \cite{Ade:2015zua}. From this figure, it is evident that the conformally coupled DE model gives rise to a larger $\sigma_8$ in comparison with the concordance model, although when including the ACA and lensing data sets, the inferred contours overlap the \textit{Planck} $\mathrm{TT+lensing}$ shaded bands. On the other hand, the CA data set combination pushes the inferred $\Omega_m$--$\sigma_8$ contours downwards, deviating from the \textit{Planck} $\mathrm{TT+lensing}$ constraint.   

From the $\mathrm{TTEE+lensing+CA+BSHE}$ data set combination we obtain a $95\%\;\mathrm{C.L.}$ upper bound of $\alpha\!\!\!\!<\!\!\!\!0.0301$, whereas the upper bound from the $\mathrm{TTEE+lensing+CA+BSHR}$ data set combination is of $\alpha\!<\!0.0325$. When we use the ACA data set instead of the CA measurements, we obtain a $95\%\;\mathrm{C.L.}$ upper bound of $\alpha\!<\!0.0423$ with $\mathrm{TTEE+lensing+ACA+}$ $\mathrm{BSHE}$ data sets, and an upper bound of $\alpha<0.0467$ with $\mathrm{TTEE+lensing+ACA+BSHR}$ data sets. Moreover, when using the $H_0^{\mathrm{E}}$ local value of the Hubble constant together with the cluster abundance data sets, a larger upper bound on $\lambda$ is allowed, in comparison with the analyses which use $H_0^{\mathrm{R}}$. We find that there is a marginal inverse correlation between $\lambda$ and $\sigma_8$, and a correlation between $H_0$ and $\sigma_8$, thus explaining these shifts in the upper bounds of $\lambda$. This is clearly illustrated in Fig. \ref{fig:alpha_lambda_ER}. Unfortunately, this relationship between $H_0$ and $\sigma_8$ would not be able to alleviate the tension between the low--redshift and high--redshift probes.

\begin{figure}
\centering
  \includegraphics[width=1.02\columnwidth]{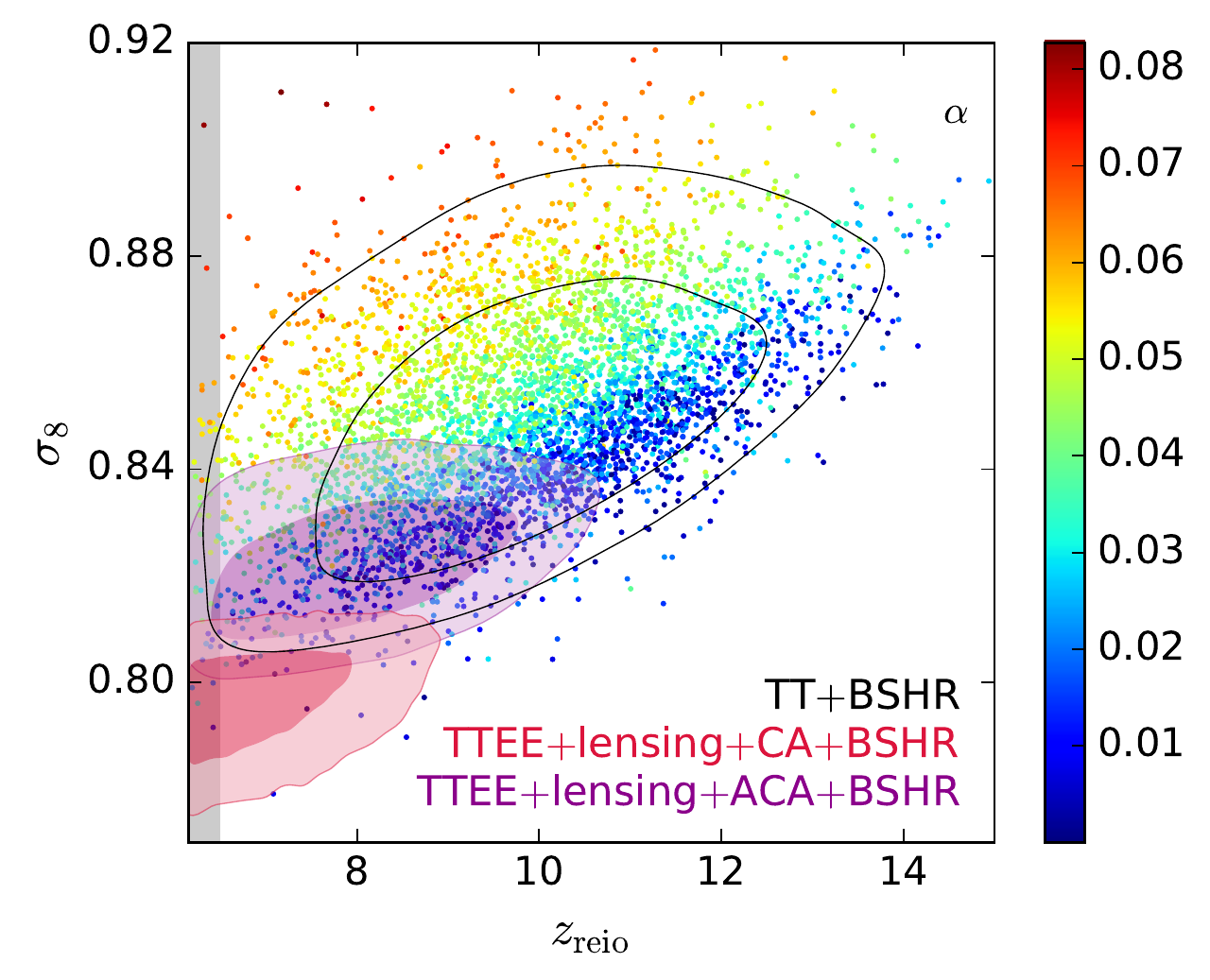}
\caption{Marginalized two--dimensional constraints on the redshift of reionization $z_{\mathrm{reio}}$ and $\sigma_8$, together with samples from the $\mathrm{TT+BSHR}$ data sets colour coded with the value of the conformal coupling parameter $\alpha$. The gray band denotes the excluded region by observations of the spectra of high redshift quasars \cite{Fan:2005es}.}  
\label{fig:conf_reio}
\end{figure}

Following the discussion on the optical depth of reionization parameter in section \ref{sec:results}, in Fig. \ref{fig:conf_reio} we show the correlation between the redshift of reionization $z_{\mathrm{reio}}$, and $\sigma_8$ in the conformally coupled model. We should remark that this relationship between the mentioned parameters also follows in the other coupled DE models. Apart from the marginalized contours from distinct data set combinations, we also include a few samples from the $\mathrm{TT+BSHR}$ data set combination colour coded with the value of $\alpha$. The marginalized contours of the $\mathrm{TT+BSHR}$ and the $\mathrm{TTEE+lensing+ACA+BSHR}$ data set combinations only overlap in a region of compatible $\sigma_8$ values with the ACA data set. Consequently, tighter constraints are placed on the $z_{\mathrm{reio}}$--$\sigma_8$--$\alpha$ subspace, placing a lower upper bound on $\alpha$. Moreover, there is a further reduction of the overlapping region between the marginalized contours of the $\mathrm{TT+BSHR}$ and the $\mathrm{TTEE+lensing+CA+BSHR}$ data set combinations, and the contour from the latter data set combination shifts downwards due to the incompatibility of the CA measurements with high $\sigma_8$ values. In Fig. \ref{fig:conf_reio}, we also show an excluded region of $z_{\mathrm{reio}}$ inferred by observations of the Gunn--Peterson effect \cite{Gunn:1965hd} in quasar spectra \cite{Fan:2005es}. As clearly shown in this figure, our constraints are in agreement with the latter observations, although a preference towards lower $\sigma_8$ values could eventually shift the marginalized contours into the excluded region. 

{\setlength\extrarowheight{5pt}
\begin{table*}
\begin{center}
\begin{tabular}{ l  c  c  c  c  c  c} 
 \hline
\hline
Parameter~  &  ~TT~ & ~$\mathrm{TT+BSHE}$~ & ~$\mathrm{TT+BSHR}$~ & ~TTEE~ & ~$\mathrm{TTEE+BSHE}$~ & ~$\mathrm{TTEE+BSHR}$~  \\
\hline
100 $\Omega_b h^2$\dotfill & $2.2288^{+0.024844}_{-0.025103}$ & $2.2318^{+0.021065}_{-0.021020}$ & $2.2400^{+0.020856}_{-0.021365}$ & $2.2279^{+0.016819}_{-0.017280}$ & $2.2321^{+0.014873}_{-0.014770}$ & $2.2380^{+0.014861}_{-0.014998}$ \\
$\Omega_c h^2$\dotfill & $0.12293^{+0.0029816}_{-0.0053997}$ & $0.12238^{+0.0021302}_{-0.0046422}$ & $0.12096^{+0.0019788}_{-0.0040800}$ & $0.12333^{+0.0023511}_{-0.0050921}$ & $0.12265^{+0.0019685}_{-0.0044646}$ & $0.12131^{+0.0018842}_{-0.0036619}$ \\
100 $\theta_s$\dotfill & $1.0419^{+0.00048030}_{-0.00047326}$ & $1.0420^{+0.00044038}_{-0.00044199}$ & $1.0421^{+0.00042639}_{-0.00043185}$ & $1.0418^{+0.00032702}_{-0.00032754}$ & $1.0419^{+0.00030542}_{-0.00030525}$ & $1.0419^{+0.00031146}_{-0.00030467}$ \\
$\tau_{\mathrm{reio}}$\dotfill & $0.078477^{+0.017682}_{-0.020519}$ & $0.079563^{+0.017484}_{-0.018663}$ & $0.082882^{+0.018012}_{-0.018692}$ & $0.077320^{+0.016929}_{-0.017617}$ & $0.079307^{+0.016578}_{-0.016441}$ & $0.082658^{+0.017012}_{-0.016557}$ \\
$\ln(10^{10}A_s)$\dotfill & $3.0895^{+0.034340}_{-0.039124}$ & $3.0912^{+0.034352}_{-0.037227}$ & $3.0958^{+0.035787}_{-0.037204}$ & $3.0891^{+0.032957}_{-0.034681}$ & $3.0919^{+0.032702}_{-0.032621}$ & $3.0973^{+0.033686}_{-0.033500}$ \\
$n_s$\dotfill & $0.96643^{+0.0065522}_{-0.0069298}$ & $0.96730^{+0.0047481}_{-0.0048034}$ & $0.96938^{+0.0046875}_{-0.0047143}$ & $0.96456^{+0.0050324}_{-0.0052587}$ & $0.96585^{+0.0043671}_{-0.0045099}$ & $0.96756^{+0.0042655}_{-0.0043602}$ \\ 
$\lambda$\dotfill & $-$ & $-$ & $-$ & $-$ & $-$ & $-$ \\ 
$D_{M}/\,\mathrm{meV}^{-1}$\ldots & $>0.4627$ & $>0.5883$ & $>0.6540$ & $>0.4599$ & $>0.6031$ & $>0.6810$ \\
\hline
$H_0$\dotfill & $67.306^{+1.8214}_{-1.0699}$ & $67.969^{+0.73556}_{-0.67588}$ & $68.510^{+0.62353}_{-0.60045}$ & $67.084^{+1.5691}_{-0.7430}$ & $67.846^{+0.66245}_{-0.53900}$ & $68.288^{+0.54432}_{-0.51848}$ \\
$\Omega_m$\dotfill & $0.32125^{+0.015937}_{-0.025371}$ & $0.31336^{+0.009237}_{-0.014280}$ & $0.30554^{+0.008590}_{-0.011918}$ & $0.32407^{+0.012147}_{-0.023283}$ & $0.31507^{+0.008146}_{-0.013818}$ & $0.30821^{+0.007811}_{-0.010956}$ \\
$\sigma_8$\dotfill & $0.90433^{+0.02913}_{-0.11800}$ & $0.92196^{+0.03372}_{-0.11257}$ & $0.92794^{+0.04409}_{-0.12384}$ & $0.90220^{+0.02698}_{-0.11265}$ & $0.92580^{+0.03830}_{-0.11204}$ & $0.93049^{+0.04199}_{-0.11698}$ \\
$z_{\mathrm{reio}}$\dotfill & $9.9392^{+1.7144}_{-1.6766}$ & $10.031^{+1.6446}_{-1.5612}$ & $10.287^{+1.6910}_{-1.5378}$ & $9.8610^{+1.6225}_{-1.4832}$ & $10.023^{+1.5982}_{-1.3557}$ & $10.292^{+1.5777}_{-1.3897}$ \\
$H_0 t_0$\dotfill & $0.94836^{+0.023160}_{-0.012187}$ & $0.95684^{+0.0093730}_{-0.0076823}$ & $0.96326^{+0.0074715}_{-0.0069426}$ & $0.94578^{+0.019865}_{-0.008976}$ & $0.95546^{+0.0084970}_{-0.0062639}$ & $0.96082^{+0.0065600}_{-0.0060206}$ \\
\hline
\hline
\end{tabular}
\end{center}
\caption{\label{table:disf_const_Tab1} For each data set combination we report the mean values and $1\sigma$ errors in the constant disformally coupled DE scenario, in which we set $\beta=0$. The Hubble constant is given in units of $\mathrm{km}\,\mathrm{s}^{-1}\,\mathrm{Mpc}^{-1}$. These data sets were not able to constrain the parameter $\lambda$.}
\end{table*}}
%

\subsection{Disformal model constraints}
\label{sec:disformal_results}
%
In this section we present and discuss the MCMC inferred parameter constraints in the constant disformally coupled DE model with the coupling parameter $D_M$, together with the exponential disformally coupled DE model with the coupling parameter $\beta$, as defined in Eq. (\ref{coupling_choice}). We will start with the former case, in which we set $\beta=0$ in the disformal coupling function $D(\phi)$, and fix the conformal coupling function to unity. In Tables \ref{table:disf_const_Tab1} and \ref{table:disf_const_Tab2} we show the parameter constraints from several data set combinations. 

Similar to the conformally coupled scenario discussed in section \ref{sec:conformal_results}, marginally tighter constraints on the varied cosmological parameters are obtained with the TTEE CMB likelihood in comparison with the TT likelihood, as clearly seen in Fig. \ref{fig:dot_plot}. In Table \ref{table:disf_const_Tab1} we present the parameter constraints inferred from the CMB likelihoods, together with the joint combination of the CMB likelihoods with the background data sets.  Although the TTEE likelihood seems to improve the parameter constraints, it is still not able to put tight constraints on the scalar field's potential parameter $\lambda$, even when this is combined with the background data sets. 

\begin{figure}
\centering
  \includegraphics[width=0.98\columnwidth]{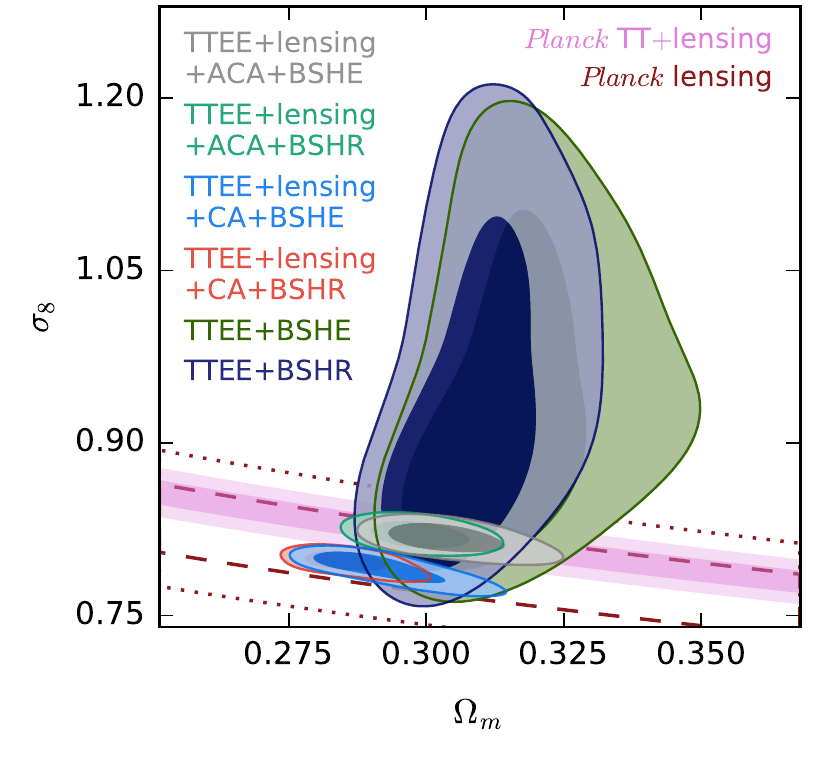}
\caption{Marginalized two--dimensional constraints on parameters of the constant disformal model using the data sets indicated in the figure. The shaded band depicts the \textit{Planck} $\mathrm{TT+lensing}$ constraint, whereas the region enclosed by the dashed $(1\sigma)$ and dotted $(2\sigma)$ lines shows the constraint from \textit{Planck} lensing alone \cite{Ade:2015zua}.}  
\label{fig:disf_const_s8_Om_bands}
\end{figure}
{\setlength\extrarowheight{5pt}
\begin{table*}
\begin{center}
\begin{tabular}{ l  c  c  c  c } 
 \hline
\hline
Parameter~  & ~\begin{tabular}[t]{@{}c@{}}$\mathrm{TTEE+lensing}$ \\ $\mathrm{+\,CA+BSHE}$\end{tabular}~ 
			& ~\begin{tabular}[t]{@{}c@{}}$\mathrm{TTEE+lensing}$ \\ $\mathrm{+\,CA+BSHR}$\end{tabular}~ 
			& ~\begin{tabular}[t]{@{}c@{}}$\mathrm{TTEE+lensing}$ \\ $\mathrm{+\,ACA+BSHE}$\end{tabular}~
			& ~\begin{tabular}[t]{@{}c@{}}$\mathrm{TTEE+lensing}$ \\ $\mathrm{+\,ACA+BSHR}$\end{tabular}~ \\ 
\hline
100 $\Omega_b h^2$\dotfill & $2.2556^{+0.014086}_{-0.014301}$ & $2.2574^{+0.013880}_{-0.013855}$ & $2.2344^{+0.014274}_{-0.014765}$ & $2.2389^{+0.014533}_{-0.014663}$ \\
$\Omega_c h^2$\dotfill & $0.11553^{+0.00086684}_{-0.00078545}$ & $0.11540^{+0.00082727}_{-0.00074345}$ & $0.11850^{+0.0010896}_{-0.0011557}$ & $0.11802^{+0.0010327}_{-0.0010857}$ \\
100 $\theta_s$\dotfill & $1.0419^{+0.00030272}_{-0.00029673}$ & $1.0420^{+0.00030035}_{-0.00029509}$ & $1.0419^{+0.00029949}_{-0.00029791}$ & $1.0419^{+0.00029929}_{-0.00030371}$ \\
$\tau_{\mathrm{reio}}$\dotfill & $0.048565^{+0.0021974}_{-0.0085645}$ & $0.048860^{+0.0023427}_{-0.0088598}$ & $0.058265^{+0.008664}_{-0.012685}$ & $0.060940^{+0.010041}_{-0.012200}$ \\
$\ln(10^{10}A_s)$\dotfill & $3.0195^{+0.009425}_{-0.016164}$ & $3.0201^{+0.010018}_{-0.016289}$ & $3.0466^{+0.016963}_{-0.022872}$ & $3.0511^{+0.019062}_{-0.022549}$ \\
$n_s$\dotfill & $0.97181^{+0.0037701}_{-0.0039588}$ & $0.97212^{+0.0037568}_{-0.0039197}$ & $0.96674^{+0.0041215}_{-0.0043175}$ & $0.96795^{+0.0041353}_{-0.0042950}$ \\ 
$\lambda$\dotfill & $<0.6720(0.9830)$ & $<0.3587(0.7270)$ & $<0.4818(0.8953)$ & $<0.3109(0.6412)$ \\ 
$D_{M}/\,\mathrm{meV}^{-1}$\ldots & $<0.2500$ & $<0.3680$ & $<0.4420$ & $<0.5730$ \\
\hline
$H_0$\dotfill & $68.817^{+1.1615}_{-0.6336}$ & $69.485^{+0.63847}_{-0.43839}$ & $68.058^{+0.80975}_{-0.55459}$ & $68.552^{+0.55946}_{-0.51014}$ \\
$\Omega_m$\dotfill & $0.29175^{+0.006038}_{-0.010323}$ & $0.28584^{+0.0048884}_{-0.0061844}$ & $0.30420^{+0.0065834}_{-0.0087393}$ & $0.29885^{+0.0061959}_{-0.0066217}$ \\
$\sigma_8$\dotfill & $0.79149^{+0.011428}_{-0.007244}$ & $0.79668^{+0.0068600}_{-0.0060705}$ & $0.81655^{+0.0094827}_{-0.0086699}$ & $0.82002^{+0.0080772}_{-0.0087196}$ \\
$z_{\mathrm{reio}}$\dotfill & $6.9709^{+0.24343}_{-0.96367}$ & $6.9948^{+0.25641}_{-0.98971}$ & $8.0308^{+0.9372}_{-1.1699}$ & $8.2758^{+1.0452}_{-1.0987}$ \\
$H_0 t_0$\dotfill & $0.96655^{+0.015143}_{-0.008064}$ & $0.97511^{+0.0085264}_{-0.0048174}$ & $0.95809^{+0.010678}_{-0.006486}$ & $0.96419^{+0.0069586}_{-0.0058817}$ \\
\hline
\hline
\end{tabular}
\end{center}
\caption{\label{table:disf_const_Tab2} For each data set combination we report the mean values and $1\sigma$ errors in the constant disformally coupled DE scenario, in which we set $\beta=0$. The Hubble constant is given in units of $\mathrm{km}\,\mathrm{s}^{-1}\,\mathrm{Mpc}^{-1}$. For the parameter $\lambda$, we also write in brackets the $2\sigma$ upper limits.}
\end{table*}}
\begin{figure}
\centering
  \includegraphics[width=1.03\columnwidth]{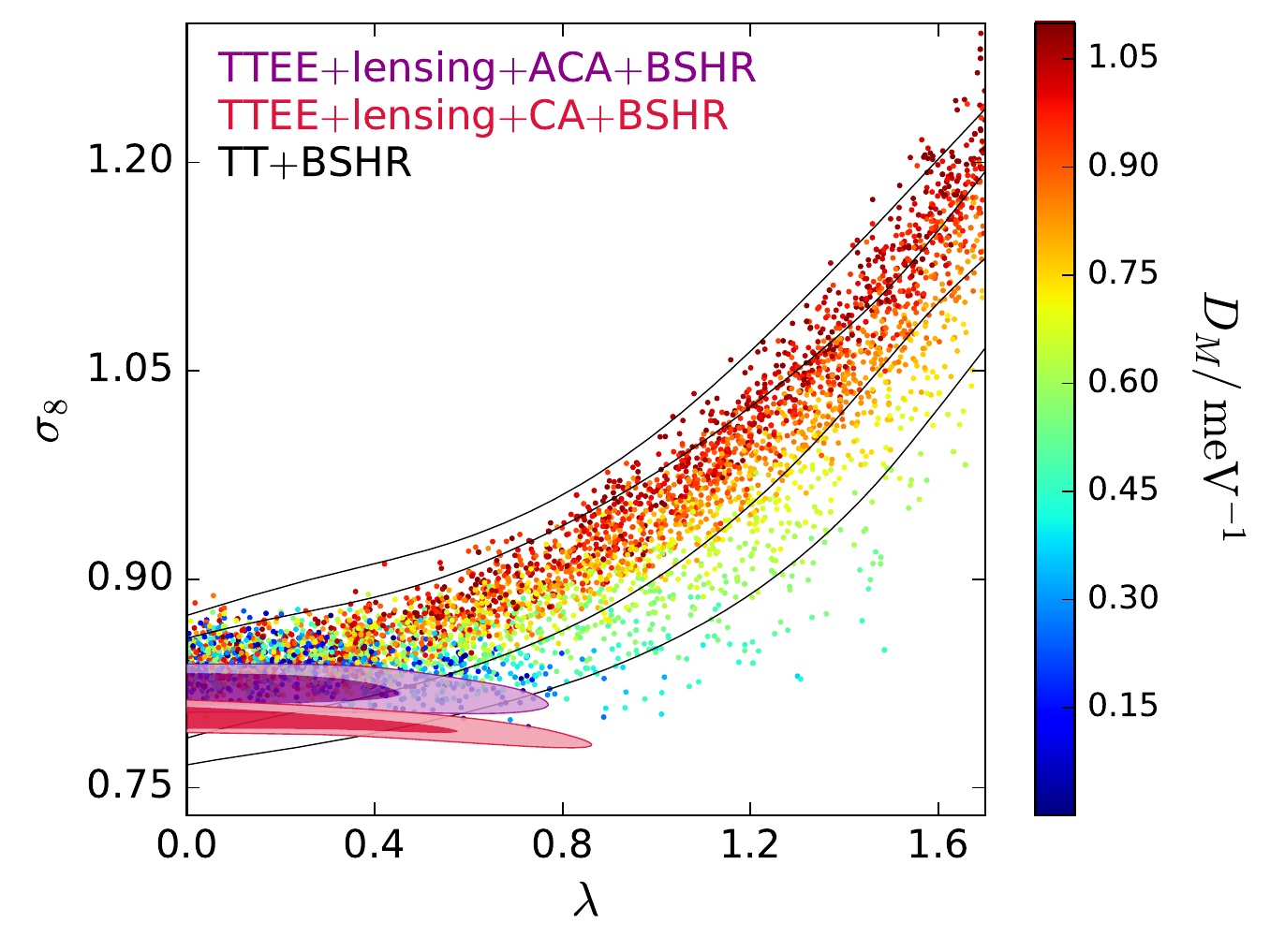}
\caption{Marginalized constraints on parameters of the constant disformal model using the data sets indicated in the figure. The sample points are taken from the $\mathrm{TT+BSHR}$ data sets and colour coded with the value of the disformal coupling parameter $D_M/\,\mathrm{meV}^{-1}$.}  
\label{fig:disf_const_s8_lambda}
\end{figure}

A striking difference between the derived cosmological parameter constraints in the conformally coupled DE model and the constant disformally coupled DE model, is the anomalous enhancement in the mean value of $\sigma_8$ in the latter coupled DE model. Other interacting DE models that are characterised with relatively high values of $\sigma_8$ were discussed in Refs. \cite{DiValentino:2017iww,Murgia:2016ccp}. Although we are considering the constant disformally coupled DE model, these features are also present in the exponential disformally coupled DE model discussed in the last part of this section. This increase in the mean value of $\sigma_8$ in the constant disformally coupled DE model is expected due to the energy transfer taking place between DM and DE, and particularly as a result of a coupling induced additional force acting between the DM particles \cite{Mifsud:2017fsy,Jack}. This fifth--force is also present in the conformally coupled DE model, although the strength of this force is found to be the largest in coupled DE models which make use of the disformal coupling. Consequently, this leads to an enhancement in the growth of perturbations in comparison with the uncoupled quintessence and conformally coupled DE models. Moreover, when a disformal coupling between DE and DM is present, this induces intermediate--scales time--dependent damped oscillations in the matter growth rate function \cite{Mifsud:2017fsy}. In these analyses we are not able to probe these scale--dependent features in the matter growth rate function, although we believe that deriving constraints from the scale--dependence of the matter growth rate function in these coupled DE models would lead to an interesting study.

The considerably large range of allowed values of $\sigma_8$ by the CMB likelihoods and the background data sets is clearly illustrated in Fig. \ref{fig:disf_const_s8_Om_bands}. The influence of the local value of the Hubble constant is mainly attributed with the constraint on $\Omega_m$, due to the inverse correlation between $\Omega_m$ and $H_0$. Although the marginalized contours of the $\mathrm{TTEE+BSHE}$ and $\mathrm{TTEE+BSHR}$ data set combinations are still in agreement with the concordance model $1\sigma$ and $2\sigma$ approximate fit constraints, very weak constraints are inferred from these data set combinations considered in Table \ref{table:disf_const_Tab1}. 

\begin{figure*}
\centering
  \includegraphics[width=0.91\columnwidth]{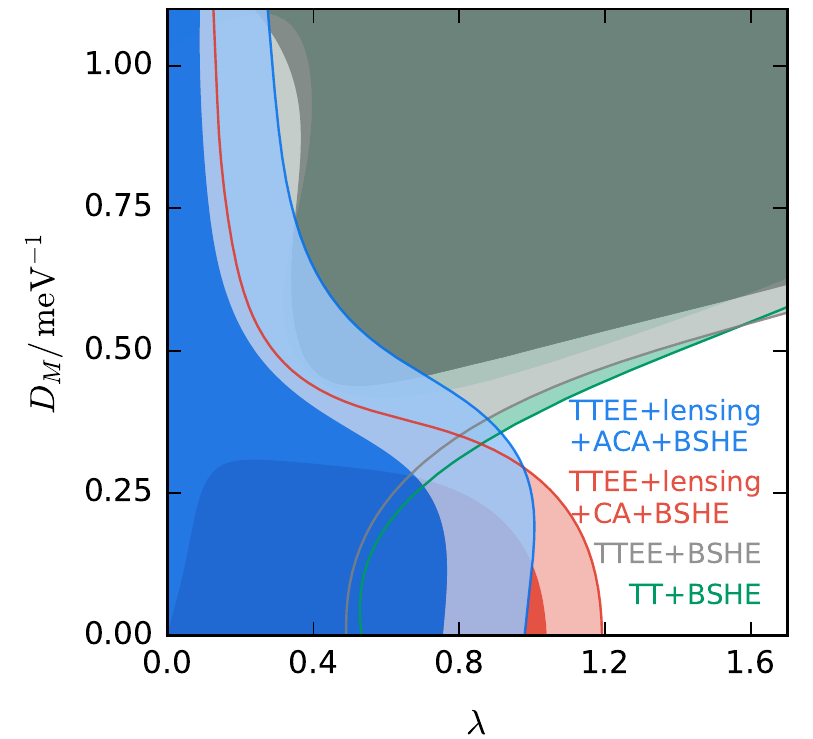}
  \includegraphics[width=1.11\columnwidth]{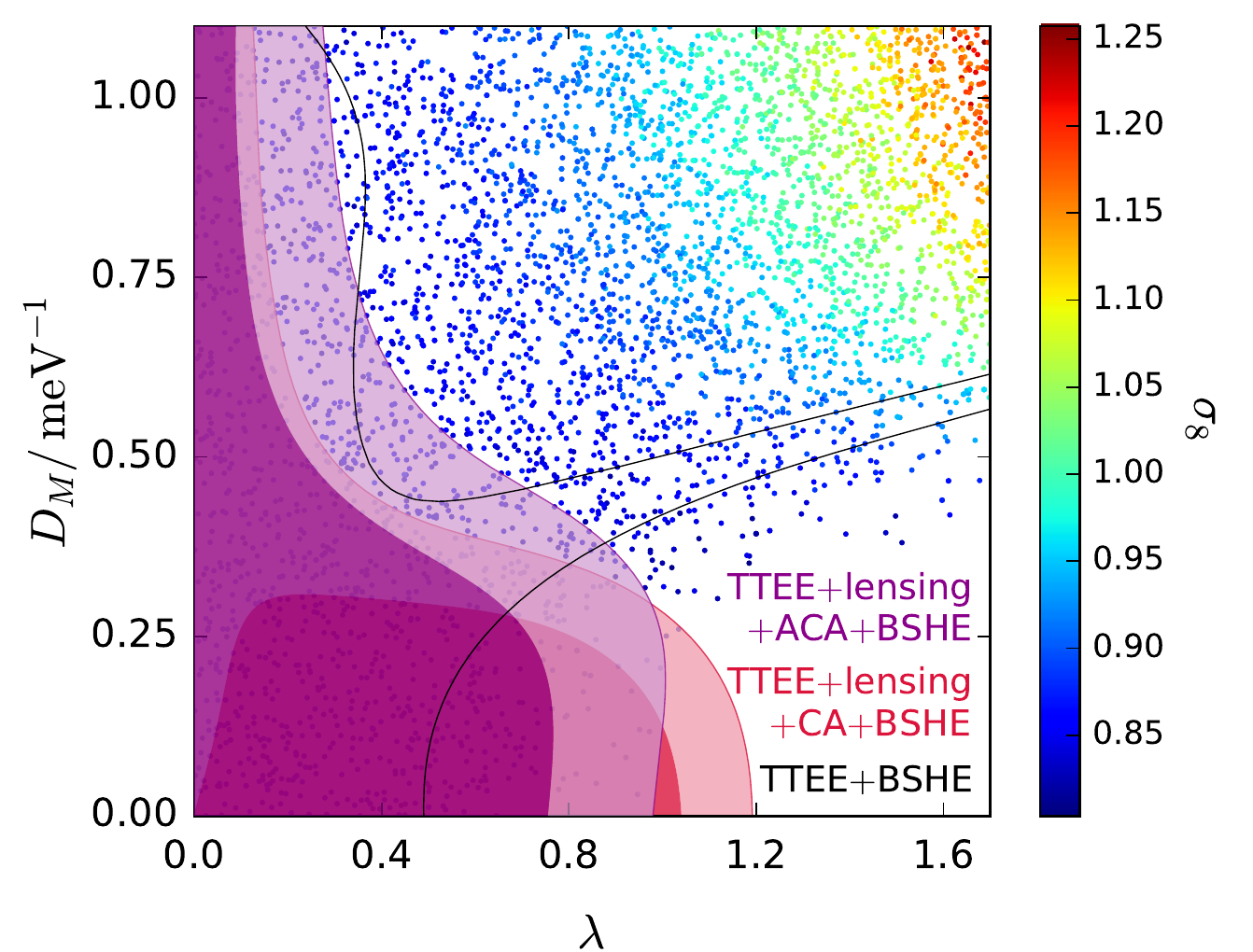}
  \includegraphics[width=0.91\columnwidth]{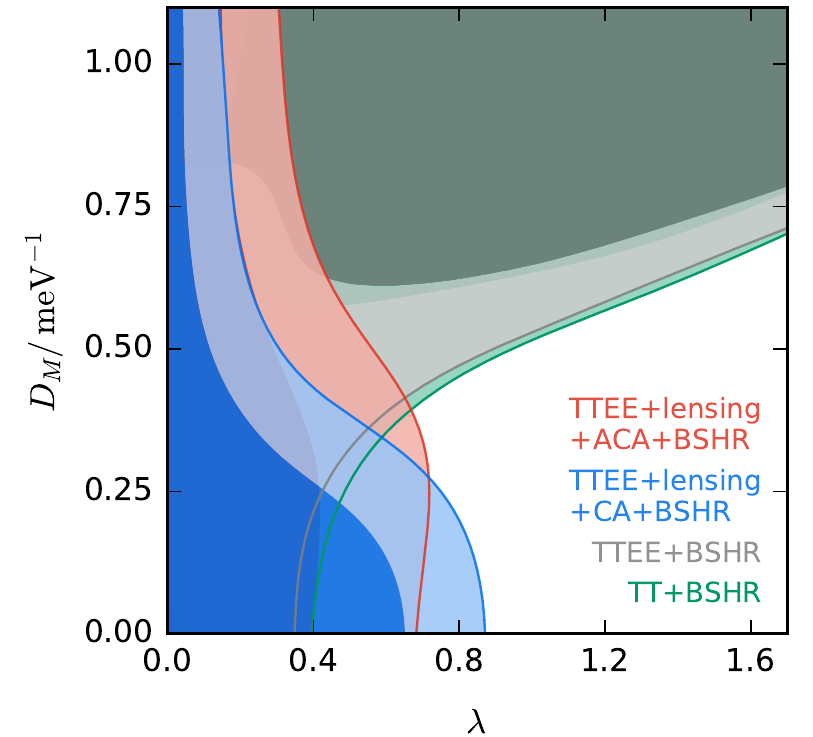}
  \includegraphics[width=1.11\columnwidth]{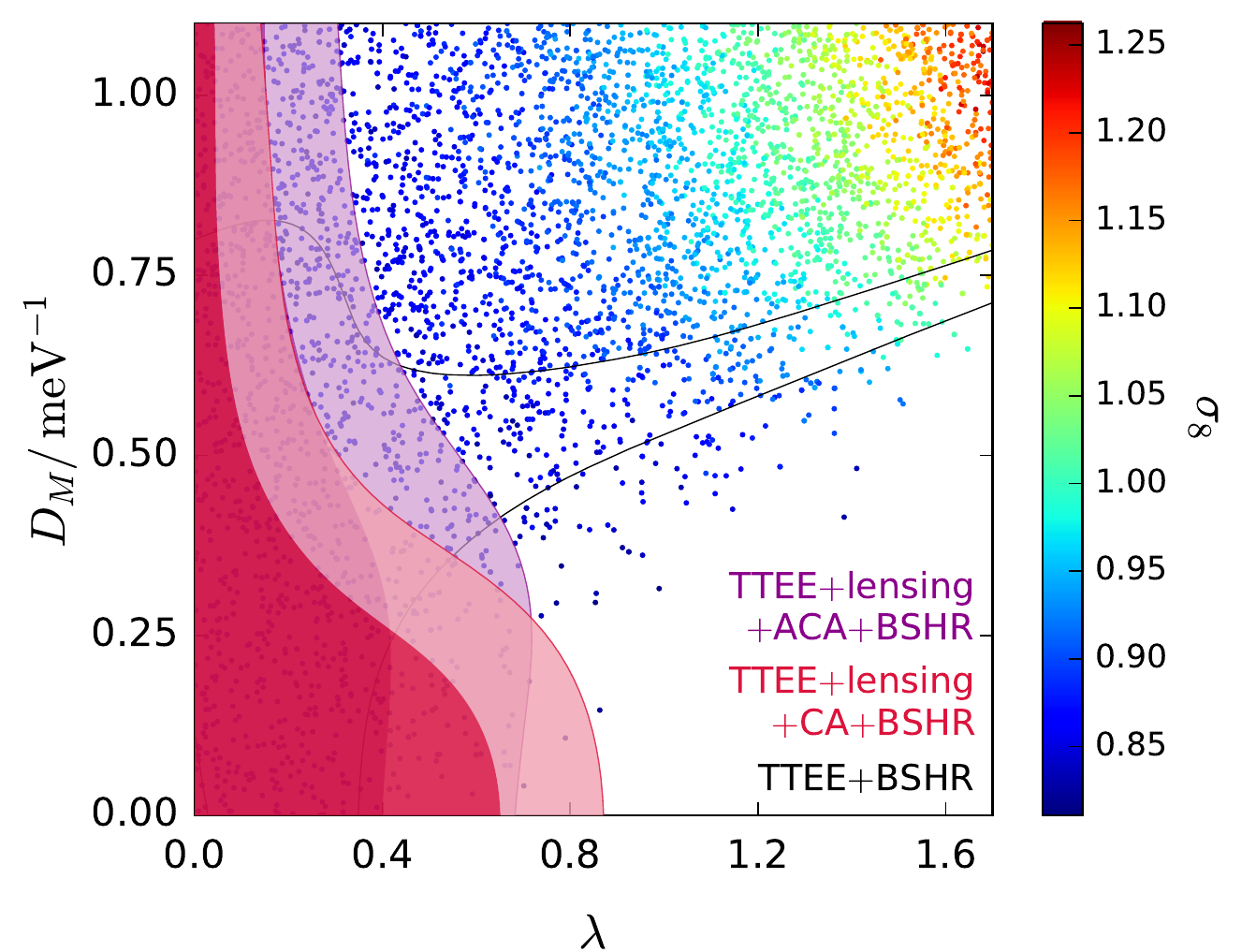}
\caption{Marginalized constraints on the disformal coupling parameter $D_M/\,\mathrm{meV}^{-1}$, and the slope of the exponential scalar field potential $\lambda$, in the constant disformally coupled DE model. In the upper two panels we use the $H_0^{\mathrm{E}}$ local value of the Hubble constant, and in the lower two panels we use $H_0^{\mathrm{R}}$. The sample points in the top right and lower right panels are colour coded with the value of $\sigma_8$, and are taken from the data sets represented by the black solid contour lines of each panel.}  
\label{fig:DM_lambda_ER}
\end{figure*}

In order to shrink these contours, we further add the cluster abundance data sets along with the CMB lensing likelihood. The inferred parameter constraints are tabulated in Table \ref{table:disf_const_Tab2}, in which we are now able to constrain the scalar field's exponent parameter $\lambda$, as clearly depicted in Fig. \ref{fig:disf_const_s8_lambda}. As expected, the measurements of the ACA and CA data sets do not allow for such large values of $\sigma_8$, and consequently shrink the marginalized contours of Fig. \ref{fig:disf_const_s8_Om_bands}. Indeed, the marginalized contours of the additional CMB lensing and ACA data sets overlap the \textit{Planck} $\mathrm{TT+lensing}$ constraint bands, similar to what happened in the conformally coupled model. Even in this model, the CA data set is still able to lower the mean value of $\sigma_8$, in order to be compatible with the relatively low $\sigma_8$ measurements of this data set. Moreover, from the coloured samples of the $\lambda$--$\sigma_8$--$D_M$ subspace of Fig. \ref{fig:disf_const_s8_lambda}, we observe that the inclusion of the cluster abundance data sets and the CMB lensing likelihood exclude the relatively high $D_M$ values which lie along the top section of the $\lambda$--$\sigma_8$ band inferred from the $\mathrm{TT+BSHR}$ data set combination. Consequently, an upper bound is placed on the disformal coupling parameter $D_M$, instead of a lower bound as reported in Table \ref{table:disf_const_Tab1}. Tight $95\%\;\mathrm{C.L.}$ upper bounds are placed on the scalar field's potential exponent parameter of $\lambda<0.9830\;\mathrm{(TTEE+lensing+CA+BSHE)}$, $\lambda<0.7270\;\mathrm{(TTEE+lensing+CA+BSHR)}$,\hspace*{0.5pt} $\lambda<0.8953\;\mathrm{(TTEE+lensing+ACA+BSHE)}$, and of\hspace*{0.5pt} $\lambda<0.6412$ $\mathrm{(TTEE+lensing+ACA+BSHR)}$. Despite\hspace*{0.5pt} of the improved constraints on the parameters, we only obtain $68\%\;\mathrm{C.L.}$ upper bounds on the constant disformal coupling parameter of $D_M<0.2500\;\mathrm{meV}^{-1}$ $\mathrm{(TTEE+lensing+CA+BSHE)}$, $D_M<0.3680\;\mathrm{meV}^{-1}$ $\mathrm{(TTEE+lensing+CA+BSHR)}$, $D_M<0.4420\;\mathrm{meV}^{-1}$ $\mathrm{(TTEE+lensing+ACA+BSHE)}$,\hspace*{1pt} and also of $D_M<0.5730\;\mathrm{meV}^{-1}$ $\mathrm{(TTEE+lensing+ACA+BSHR)}$.  
{\setlength\extrarowheight{5pt}
\begin{table*}
\begin{center}
\begin{tabular}{ l  c  c  c  c } 
 \hline
\hline
Parameter~  & ~\begin{tabular}[t]{@{}c@{}}$\mathrm{TTEE+lensing}$ \\ $\mathrm{+\,CA+BSHE}$\end{tabular}~ 
			& ~\begin{tabular}[t]{@{}c@{}}$\mathrm{TTEE+lensing}$ \\ $\mathrm{+\,CA+BSHR}$\end{tabular}~ 
			& ~\begin{tabular}[t]{@{}c@{}}$\mathrm{TTEE+lensing}$ \\ $\mathrm{+\,ACA+BSHE}$\end{tabular}~
			& ~\begin{tabular}[t]{@{}c@{}}$\mathrm{TTEE+lensing}$ \\ $\mathrm{+\,ACA+BSHR}$\end{tabular}~ \\ 
\hline
100 $\Omega_b h^2$\dotfill & $2.2539^{+0.013685}_{-0.013793}$ & $2.2568^{+0.013693}_{-0.013549}$ & $2.2341^{+0.014362}_{-0.014479}$ & $2.2386^{+0.014153}_{-0.014582}$ \\
$\Omega_c h^2$\dotfill & $0.11572^{+0.00077472}_{-0.00074705}$ & $0.11549^{+0.00080077}_{-0.00076624}$ & $0.11877^{+0.0010556}_{-0.0013004}$ & $0.11816^{+0.0010423}_{-0.0011337}$ \\
100 $\theta_s$\dotfill & $1.0419^{+0.00029898}_{-0.00029958}$ & $1.0420^{+0.00029930}_{-0.00029409}$ & $1.0419^{+0.00029115}_{-0.00030380}$ & $1.0419^{+0.00029922}_{-0.00029910}$ \\
$\tau_{\mathrm{reio}}$\dotfill & $0.047123^{+0.0017798}_{-0.0071226}$ & $0.048175^{+0.0021763}_{-0.0081740}$ & $0.056890^{+0.007961}_{-0.012251}$ & $0.059970^{+0.009571}_{-0.012187}$ \\
$\ln(10^{10}A_s)$\ldots & $3.0169^{+0.008201}_{-0.014090}$ & $3.0188^{+0.009080}_{-0.015124}$ & $3.0440^{+0.016027}_{-0.021780}$ & $3.0492^{+0.018101}_{-0.022419}$ \\
$n_s$\dotfill & $0.97118^{+0.0037534}_{-0.0038919}$ & $0.97187^{+0.0037082}_{-0.0038410}$ & $0.96651^{+0.0040552}_{-0.0042174}$ & $0.96773^{+0.0040610}_{-0.0042276}$ \\ 
$\lambda$\dotfill & $<0.1269(0.3360)$ & $<0.1257(0.3090)$ & $<0.2841(0.7847)$ & $<0.2458(0.5702)$ \\ 
$\beta$\dotfill & $<1.6700$ & $<1.7100$ & $<1.5900$ & $<1.7412$ \\
\hline
$H_0$\dotfill & $69.653^{+0.36892}_{-0.37517}$ & $69.783^{+0.35347}_{-0.38646}$ & $68.356^{+0.59883}_{-0.47534}$ & $68.658^{+0.48105}_{-0.48065}$ \\
$\Omega_m$\dotfill & $0.28502^{+0.0043860}_{-0.0043519}$ & $0.28354^{+0.0044259}_{-0.0042951}$ & $0.30209^{+0.0060342}_{-0.0077797}$ & $0.29820^{+0.0060958}_{-0.0063387}$ \\
$\sigma_8$\dotfill & $0.79966^{+0.0039381}_{-0.0050455}$ & $0.79975^{+0.0041628}_{-0.0053221}$ & $0.82041^{+0.0080360}_{-0.0082640}$ & $0.82179^{+0.0074551}_{-0.0086093}$ \\
$z_{\mathrm{reio}}$\dotfill & $6.8285^{+0.19966}_{-0.81636}$ & $6.9273^{+0.23762}_{-0.91510}$ & $7.8974^{+0.8731}_{-1.1287}$ & $8.1825^{+0.9975}_{-1.1125}$ \\
$H_0 t_0$\dotfill & $0.97764^{+0.0040606}_{-0.0044758}$ & $0.97906^{+0.0040769}_{-0.0043334}$ & $0.96203^{+0.0069251}_{-0.0059476}$ & $0.96562^{+0.0056395}_{-0.0056261}$ \\
\hline
\hline
\end{tabular}
\end{center}
\caption{\label{table:disf_exp_Tab} For each data set combination we report the mean values and $1\sigma$ errors in the exponential disformally coupled DE scenario, in which we set $D_M V_0=1$. The Hubble constant is given in units of $\mathrm{km}\,\mathrm{s}^{-1}\,\mathrm{Mpc}^{-1}$. For the model parameter $\lambda$, we also write in brackets the $2\sigma$ upper limits. }
\end{table*}}
\begin{figure*}
\centering
  \includegraphics[width=0.98\columnwidth]{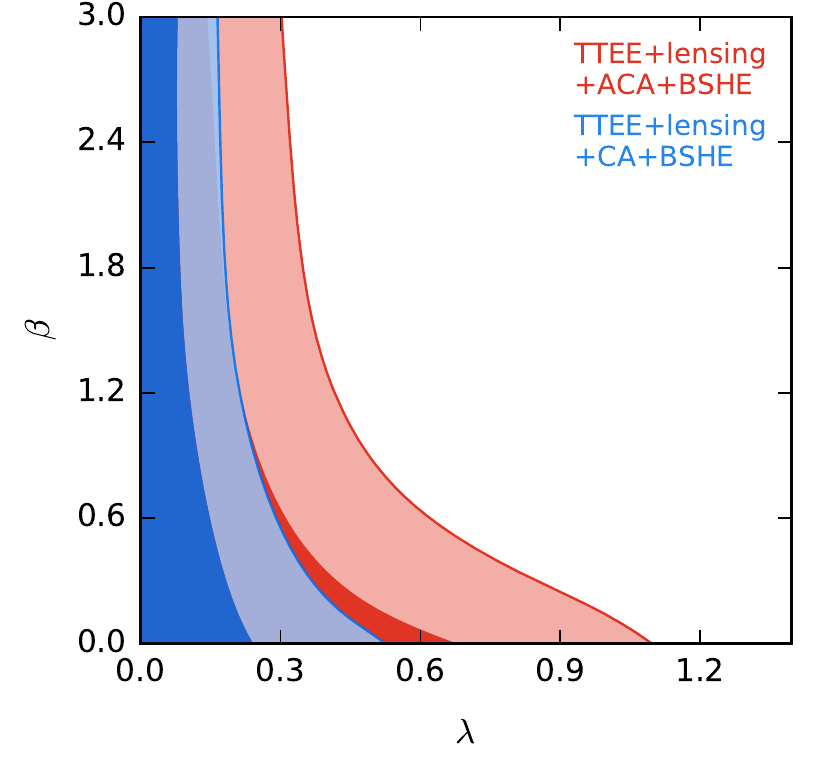}
  \includegraphics[width=0.98\columnwidth]{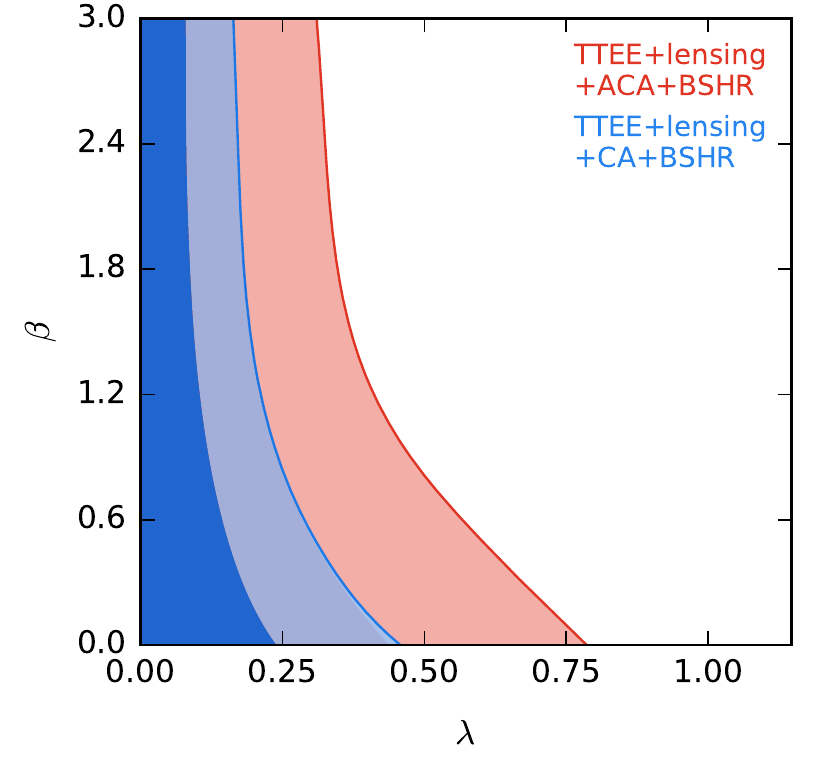}
\caption{Marginalized two--dimensional constraints on the exponential disformal coupling parameter $\beta$, and the slope of the exponential potential $\lambda$. In the left panel we use the $H_0^{\mathrm{E}}$ local value of the Hubble constant, and in the right panel we use $H_0^{\mathrm{R}}$. In this model we set $D_M V_0=1$.}  
\label{fig:beta_lambda_ER}
\end{figure*}
\\\hspace*{1em}In the upper left and lower left panels of Fig. \ref{fig:DM_lambda_ER}, we compare the two--dimensional marginalized constraints on the constant disformally coupled DE model parameters $D_M$ and $\lambda$. We complement these marginalized constraints by their respective $\lambda$--$D_M$--$\sigma_8$ subspace, which we show on the right hand side of these panels. Undoubtedly, the allowed large values of $D_M$ and $\lambda$ by the CMB likelihoods together with the background data sets, will be excluded by the cluster abundance data sets. This is evidently illustrated by the samples located in the vicinity of the top right corner of the panels in Fig. \ref{fig:DM_lambda_ER} depicting the $\lambda$--$D_M$--$\sigma_8$ subspace. Thus, the cluster abundance data sets together with the CMB lensing likelihood are able to significantly shrink the allowed range of the parameters $D_M$ and $\lambda$. These improved constraints complement the analyses of Ref. \cite{vandeBruck:2016hpz}, in which only the background evolution was considered.
{\setlength\extrarowheight{5pt}
\begin{table*}
\begin{center}
\begin{tabular}{ l  c  c  c  c  c  c} 
 \hline
\hline
Parameter~  &  ~TT~ & ~$\mathrm{TT+BSHE}$~ & ~$\mathrm{TT+BSHR}$~ & ~TTEE~ & ~$\mathrm{TTEE+BSHE}$~ & ~$\mathrm{TTEE+BSHR}$~  \\ 
\hline
100 $\Omega_b h^2$\dotfill & $2.2292^{+0.024070}_{-0.025885}$ & $2.2308^{+0.021556}_{-0.021137}$ & $2.2378^{+0.021367}_{-0.021362}$ & $2.2282^{+0.016842}_{-0.017078}$ & $2.2309^{+0.014764}_{-0.014974}$ & $2.2358^{+0.014823}_{-0.014863}$ \\
$\Omega_c h^2$\dotfill & $0.12073^{+0.0056032}_{-0.0048823}$ & $0.12124^{+0.0035890}_{-0.0042940}$ & $0.11914^{+0.0045685}_{-0.0040624}$ & $0.12121^{+0.0052433}_{-0.0044433}$ & $0.12141^{+0.0034131}_{-0.0042356}$ & $0.11947^{+0.0041912}_{-0.0036955}$ \\
100 $\theta_s$\dotfill & $1.0420^{+0.00049356}_{-0.00048926}$ & $1.0420^{+0.00043577}_{-0.00043216}$ & $1.0421^{+0.00044003}_{-0.00042633}$ & $1.0418^{+0.00032592}_{-0.00032845}$ & $1.0418^{+0.00030304}_{-0.00031205}$ & $1.0419^{+0.00029529}_{-0.00029724}$ \\
$\tau_{\mathrm{reio}}$\dotfill & $0.078386^{+0.017528}_{-0.020861}$ & $0.078803^{+0.017096}_{-0.019199}$ & $0.081653^{+0.017552}_{-0.018949}$ & $0.077251^{+0.016272}_{-0.017890}$ & $0.078334^{+0.016638}_{-0.017105}$ & $0.080978^{+0.015709}_{-0.016581}$ \\
$\ln(10^{10}A_s)$\dotfill & $3.0894^{+0.034103}_{-0.040353}$ & $3.0900^{+0.034020}_{-0.038137}$ & $3.0942^{+0.035617}_{-0.036862}$ & $3.0889^{+0.032309}_{-0.034397}$ & $3.0902^{+0.032679}_{-0.033774}$ & $3.0946^{+0.031396}_{-0.032517}$ \\
$n_s$\dotfill & $0.96641^{+0.0066510}_{-0.0068200}$ & $0.96672^{+0.0048781}_{-0.0049341}$ & $0.96851^{+0.0048974}_{-0.0048821}$ & $0.96453^{+0.0050004}_{-0.0051770}$ & $0.96556^{+0.0044388}_{-0.0044619}$ & $0.96682^{+0.0043650}_{-0.0042879}$ \\ 
$\lambda$\dotfill & $>0.7928$ & $>0.7284$ & $0.88518^{+0.64662}_{-0.41172}$ & $>0.7964$ & $>0.7329$ & $0.90123^{+0.63771}_{-0.38682}$ \\ 
$\alpha$\dotfill & $<0.3151$ & $<0.3320$ & $>0.1747$ & $<0.3127$ & $<0.3380$ & $>0.1801$ \\
$D_{M}/\,\mathrm{meV}^{-1}$\ldots & $>0.4820$ & $>0.5890(0.2578)$ & $>0.5871(0.2271)$ & $>0.4970$ & $>0.5970(0.2609)$ & $>0.5665(0.2068)$ \\
\hline
$H_0$\dotfill & $68.040^{+1.5389}_{-1.5534}$ & $68.135^{+0.70937}_{-0.65359}$ & $68.691^{+0.64172}_{-0.68003}$ & $67.811^{+1.3887}_{-1.2809}$ & $68.026^{+0.64763}_{-0.57256}$ & $68.491^{+0.53326}_{-0.61870}$ \\
$\Omega_m$\dotfill & $0.30970^{+0.022861}_{-0.022673}$ & $0.30937^{+0.012012}_{-0.013509}$ & $0.30010^{+0.013785}_{-0.012224}$ & $0.31265^{+0.020527}_{-0.018741}$ & $0.31072^{+0.011249}_{-0.013236}$ & $0.30249^{+0.013374}_{-0.010919}$ \\
$\sigma_8$\dotfill & $0.93857^{+0.05738}_{-0.10765}$ & $0.94573^{+0.04282}_{-0.11852}$ & $0.94939^{+0.04631}_{-0.11513}$ & $0.93770^{+0.04933}_{-0.10742}$ & $0.94709^{+0.04707}_{-0.11157}$ & $0.94865^{+0.04733}_{-0.10816}$ \\
$z_{\mathrm{reio}}$\dotfill & $9.9238^{+1.6998}_{-1.7209}$ & $9.9678^{+1.6604}_{-1.5971}$ & $10.190^{+1.6728}_{-1.5416}$ & $9.8484^{+1.5867}_{-1.4909}$ & $9.9381^{+1.5636}_{-1.4472}$ & $10.156^{+1.4740}_{-1.3804}$ \\
$H_0 t_0$\dotfill & $0.95735^{+0.018989}_{-0.017262}$ & $0.95894^{+0.0087876}_{-0.0079006}$ & $0.96559^{+0.0079535}_{-0.0082058}$ & $0.95471^{+0.015964}_{-0.014838}$ & $0.95775^{+0.0077619}_{-0.0071100}$ & $0.96333^{+0.0065955}_{-0.0074560}$ \\
\hline
\hline
\end{tabular}
\end{center}
\caption{\label{table:mixed_Tab1} For each data set combination we report the mean values and $1\sigma$ errors in the mixed coupled DE model. The Hubble constant is given in units of $\mathrm{km}\,\mathrm{s}^{-1}\,\mathrm{Mpc}^{-1}$. When necessary, we also write in brackets the $2\sigma$ lower limits of the model parameter $D_M$. }
\end{table*}}
\\\hspace*{1em}As we indicated in the beginning of this section, we will now consider an exponential disformally coupled model. In this case, we will still set the conformal coupling to unity, but without loss of generality we also fix the constant disformal coupling parameter to $D_M V_0=1$. Thus, in this disformally coupled DE model we vary the disformal coupling parameter $\beta$ in the MCMC analyses. Since the tightest constraints in the constant disformally coupled DE model were obtained when the cluster abundance data sets were considered in the data set combinations, we here only report and discuss the inferred parameter constraints with these data set combinations. These are tabulated in Table \ref{table:disf_exp_Tab}, and in Fig. \ref{fig:beta_lambda_ER} we illustrate the two--dimensional marginalized constraints of $\beta$ and $\lambda$. In the left panel of this figure we use the $H_0^{\mathrm{E}}$ local value of the Hubble constant, and in the right panel we instead use $H_0^{\mathrm{R}}$, in order to asses their impact on our constraints. Indeed, marginally higher upper bounds for $\lambda$ are obtained with the $H_0^{\mathrm{E}}$ Hubble constant in comparison with the inferred upper bounds from the data set combinations using $H_0^{\mathrm{R}}$. In all analyses presented in Table \ref{table:disf_exp_Tab}, we only obtain $68\%\;\mathrm{C.L.}$ upper bounds on $\beta$ which are consistent with zero. Similar to the constant disformally coupled DE model analyses, we observe that in this exponential model, the CA data set combinations put tighter constraints on the model parameters in comparison with the ACA data set combinations, especially on the exponent of the scalar field potential $\lambda$. Thus, disformally coupled DE models will be further constrained by forthcoming surveys of the large scale structures in the Universe.

\subsection{Mixed model constraints}
\label{sec:mixed_results}

In this section we discuss the derived parameter constraints in the mixed coupled DE model which simultaneously makes use of the conformal and disformal couplings. In this model we thus have an extra parameter in our MCMC analyses when compared with the number of parameters in the previous models. We will only consider a constant disformal coupling in this model, since from section \ref{sec:disformal_results} it was evident that the constant disformal model and the exponential disformal model behave in a very similar way. In Tables \ref{table:mixed_Tab1} and \ref{table:mixed_Tab2} we present the parameter constraints from several data set combinations. 

{\setlength\extrarowheight{5pt}
\begin{table*}
\begin{center}
\begin{tabular}{ l  c  c  c  c  c } 
 \hline
\hline
Parameter~  & ~\begin{tabular}[t]{@{}c@{}}$\mathrm{TTEE+lensing}$ \\ $\mathrm{+\,CA+BSHE}$\end{tabular}~ 
			& ~\begin{tabular}[t]{@{}c@{}}$\mathrm{TTEE+lensing}$ \\ $\mathrm{+\,CA+BSHR}$\end{tabular}~ 
			& ~\begin{tabular}[t]{@{}c@{}}$\mathrm{TTEE+lensing}$ \\ $\mathrm{+\,ACA+BSHE}$\end{tabular}~
			& ~\begin{tabular}[t]{@{}c@{}}$\mathrm{TTEE+lensing}$ \\ $\mathrm{+\,ACA+BSHR}$\end{tabular}~
			& ~\begin{tabular}[t]{@{}c@{}}$\mathrm{TTEE+lensing}$ \\ $\mathrm{+\,CA+BSHE}$ \\ $(D_{M}V_{0}=1)$\end{tabular}~ \\ 
\hline
100 $\Omega_b h^2$\dotfill & $2.2549^{+0.013564}_{-0.014125}$ & $2.2572^{+0.013420}_{-0.014026}$ & $2.2342^{+0.014547}_{-0.014564}$ & $2.2387^{+0.014630}_{-0.014466}$ & $2.2544^{+0.013788}_{-0.013691}$ \\
$\Omega_c h^2$\dotfill & $0.11551^{+0.00088384}_{-0.00079951}$ & $0.11523^{+0.00091116}_{-0.00077784}$ & $0.11816^{+0.0013421}_{-0.0011604}$ & $0.11753^{+0.0013973}_{-0.0011453}$ & $0.11563^{+0.00093190}_{-0.00087763}$ \\
100 $\theta_s$\dotfill & $1.0419^{+0.00029389}_{-0.00029222}$ & $1.0419^{+0.00029725}_{-0.00029499}$ & $1.0419^{+0.00030364}_{-0.00029436}$ & $1.0419^{+0.00030483}_{-0.00030791}$ & $1.0419^{+0.00029464}_{-0.00029509}$ \\
$\tau_{\mathrm{reio}}$\dotfill & $0.047592^{+0.0018186}_{-0.0075918}$ & $0.048296^{+0.0021501}_{-0.0082952}$ & $0.056709^{+0.007606}_{-0.012736}$ & $0.059471^{+0.009290}_{-0.012585}$ & $0.047212^{+0.0018168}_{-0.0072108}$ \\
$\ln(10^{10}A_s)$\dotfill & $3.0176^{+0.008268}_{-0.015084}$ & $3.0188^{+0.009592}_{-0.015573}$ & $3.0436^{+0.016019}_{-0.022268}$ & $3.0483^{+0.017962}_{-0.022847}$ & $3.0166^{+0.008266}_{-0.014585}$ \\
$n_s$\dotfill & $0.97150^{+0.0037783}_{-0.0038027}$ & $0.97198^{+0.0036723}_{-0.0039286}$ & $0.96654^{+0.0039732}_{-0.0042997}$ & $0.96764^{+0.0042007}_{-0.0042585}$ & $0.97126^{+0.0038333}_{-0.0038197}$ \\ 
$\lambda$\dotfill & $<0.3390(0.9150)$ & $<0.1420(0.5810)$ & $<0.3000(0.8530)$ & $<0.2350(0.5880)$ & $<0.1990(0.5360)$ \\ 
$\alpha$\dotfill & $<0.0879(0.3323)$ & $<0.1200(0.3541)$ & $<0.1450$ & $<0.1730$ & $<0.0522(0.1160)$ \\
$D_{M}/\,\mathrm{meV}^{-1}$\ldots & $-$ & $>0.3670$ & $>0.3590$ & $>0.4490$ & $/$ \\
\hline
$H_0$\dotfill & $69.317^{+0.86060}_{-0.35109}$ & $69.722^{+0.51967}_{-0.40274}$ & $68.329^{+0.66613}_{-0.53753}$ & $68.697^{+0.51523}_{-0.50244}$ & $69.630^{+0.45991}_{-0.40728}$ \\
$\Omega_m$\dotfill & $0.28746^{+0.0044275}_{-0.0087334}$ & $0.28353^{+0.0049905}_{-0.0052551}$ & $0.30105^{+0.0065178}_{-0.0084079}$ & $0.29655^{+0.0065002}_{-0.0065496}$ & $0.28503^{+0.0050789}_{-0.0055116}$ \\
$\sigma_8$\dotfill & $0.79682^{+0.0098000}_{-0.0050556}$ & $0.79975^{+0.0055385}_{-0.0061869}$ & $0.82170^{+0.0091970}_{-0.0097256}$ & $0.82427^{+0.008374}_{-0.010104}$ & $0.79992^{+0.0049880}_{-0.0059609}$ \\
$z_{\mathrm{reio}}$\dotfill & $6.8724^{+0.20789}_{-0.87501}$ & $6.9377^{+0.23945}_{-0.93890}$ & $7.8773^{+0.8485}_{-1.1835}$ & $8.1323^{+0.9787}_{-1.1564}$ & $6.8352^{+0.20748}_{-0.83141}$ \\
$H_0 t_0$\dotfill & $0.97307^{+0.010816}_{-0.004053}$ & $0.97821^{+0.0060367}_{-0.0047148}$ & $0.96159^{+0.0081954}_{-0.0059953}$ & $0.96605^{+0.0063939}_{-0.0056112}$ & $0.97728^{+0.0054688}_{-0.0050108}$ \\
\hline
\hline
\end{tabular}
\end{center}
\caption{\label{table:mixed_Tab2} For each data set combination we report the mean values and $1\sigma$ errors in the mixed coupled DE model. The Hubble constant is given in units of $\mathrm{km}\,\mathrm{s}^{-1}\,\mathrm{Mpc}^{-1}$. The first data set combination was not able to constrain the parameter $D_M$. In the last column, we consider the mixed coupled DE model subject to $D_M V_0=1$, thus $D_M$ is fixed in this case. When necessary, we also write in brackets the $2\sigma$ upper limits of the model parameters $\lambda$ and $\alpha$. }
\end{table*}}
\begin{figure*}
\centering
  \includegraphics[width=0.875\columnwidth]{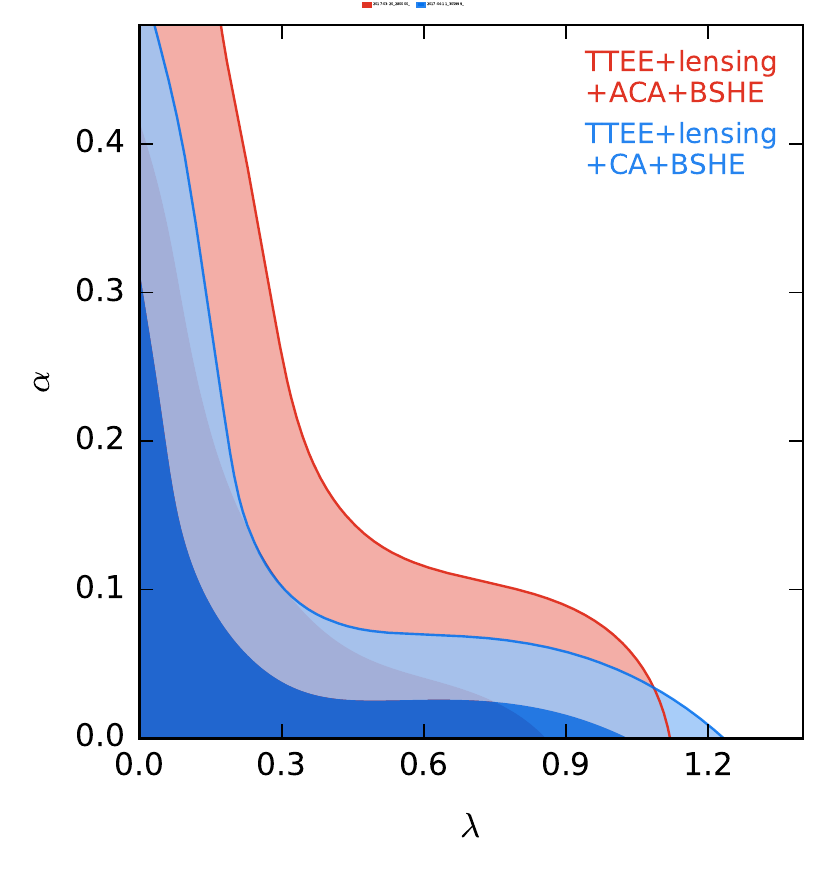}
  \includegraphics[width=0.875\columnwidth]{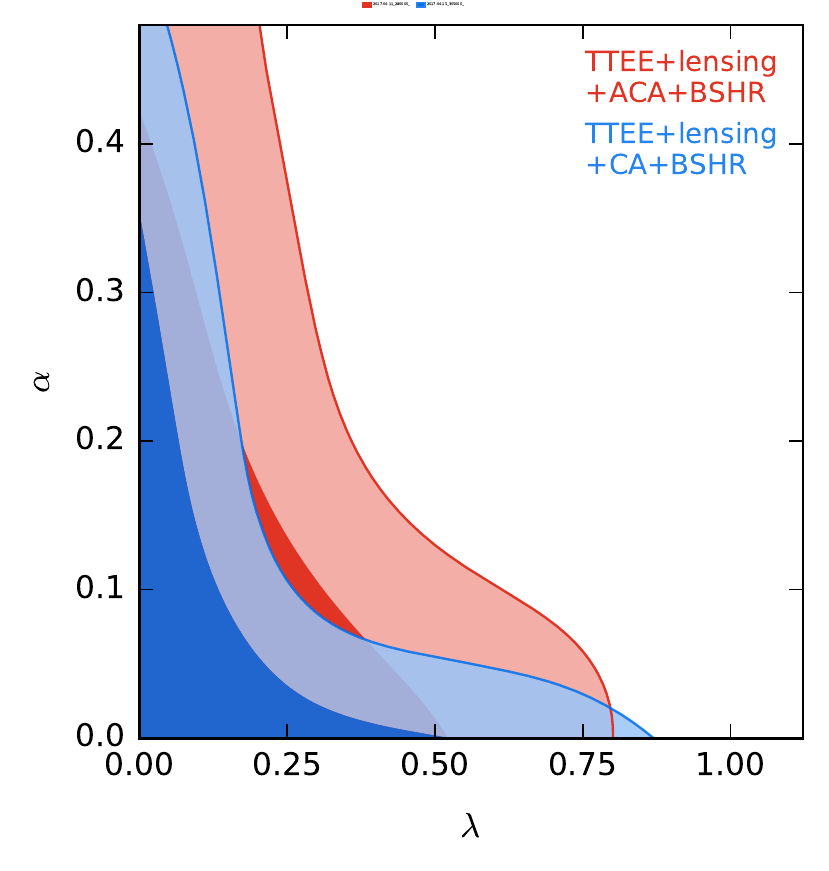}
\caption{Marginalized two--dimensional constraints on the conformal coupling parameter $\alpha$, and the slope of the exponential potential $\lambda$, in the mixed model with the parameter constraints tabulated in Table \ref{table:mixed_Tab2}. In the left panel we use the local value of the Hubble constant $H_0^{\mathrm{E}}$, whereas in the right panel we use $H_0^{\mathrm{R}}$.}  
\label{fig:alpha_lambda_mixed}
\end{figure*}

The CMB likelihoods together with the additional information from the background data sets were able to put $95\%\;\mathrm{C.L.}$ lower bounds on $D_M$, although only $68\%\;\mathrm{C.L.}$ constraints were placed on the parameters $\lambda$ and $\alpha$. The relatively high value of the Hubble constant $H_0^{\mathrm{R}}$ slightly alters the constraints on the parameters $\lambda$ and $\alpha$, in comparison with the inferred constraints with $H_0^{\mathrm{E}}$. Indeed, a significant peak in the marginalized posterior distribution of $\lambda$ is derived only in the MCMC analyses with the CMB likelihood and background data set combinations which include $H_0^{\mathrm{R}}$. Moreover, a lower bound on $\alpha$ is reported in Table \ref{table:mixed_Tab1} with the $\mathrm{TT+BSHR}$ and $\mathrm{TTEE+BSHR}$ data set combinations, instead of an upper bound which is derived from the other data sets. Furthermore, the $\mathrm{TT+BSHE}$ and $\mathrm{TTEE+BSHE}$ data set combinations prefer higher values of $D_M$ when compared with the inferred lower bounds from the $\mathrm{TT+BSHR}$ and $\mathrm{TTEE+BSHR}$ data set combinations. As clearly illustrated in Fig. \ref{fig:dot_plot}, marginally tighter constraints on the cosmological parameters are obtained with the TTEE CMB likelihood in comparison with the TT likelihood, henceforth we will only consider the TTEE CMB likelihood in the data set combinations that include the cluster abundance data sets. 

In this mixed coupled DE model, relatively high values of $\sigma_8$ are allowed by the CMB likelihoods along with the background data sets. This is expected since in this model the disformal coupling enhances the gravitational attraction between the DM particles leading to an enhancement in the growth of perturbations, in similarity with the pure disformal coupling cases discussed in section \ref{sec:disformal_results}. Therefore, in Table \ref{table:mixed_Tab2} we consider the data set combinations which are able to probe the growth of perturbations better than the CMB temperature and polarization likelihoods along with the background data sets. Indeed, significantly tighter constraints are placed on $\sigma_8$ with the additional cluster abundance data set combinations together with the CMB lensing likelihood. 

Consequently, marginally tighter constraints are derived for the mixed model parameters $\alpha$ and $\lambda$. Remarkably, the data set combinations with the BSHR background data set considered in Table \ref{table:mixed_Tab2}, were able to place upper bounds instead of lower bounds on the conformal coupling parameter $\alpha$, as reported in Table \ref{table:mixed_Tab1}. This is expected due to the correlation between $\sigma_8$ and $\alpha$. The CA data set combinations, which prefer low values of $\sigma_8$, tightly constrained the conformal coupling parameter to $\alpha<0.3323\;\mathrm{(TTEE+lensing+CA+BSHE)}$ and $\alpha<0.3541\;\mathrm{(TTEE+lensing+CA+BSHR)}$ at the $95\%$ confidence level. This is a significant improvement on the inferred constraints of Ref. \cite{vandeBruck:2016hpz}, in which only the background evolution was considered. As expected, the ACA data set combinations allow for slightly larger values of $\alpha$, since the measurements in this data set allow for marginally larger values of $\sigma_8$ which are consistent with the concordance model. The local value of the Hubble constant has a minor influence on the $95\%\;\mathrm{C.L.}$ upper bounds of $\alpha$, although the BSHR data set combinations put tighter constraints on $\lambda$ in comparison with the inferred upper bounds from the BSHE data set combinations. We show the marginalized two--dimensional constraints on $\alpha$ and $\lambda$ from the first four data set combinations of Table \ref{table:mixed_Tab2} in the panels of Fig. \ref{fig:alpha_lambda_mixed}.

\begin{figure}
\centering
  \includegraphics[width=0.98\columnwidth]{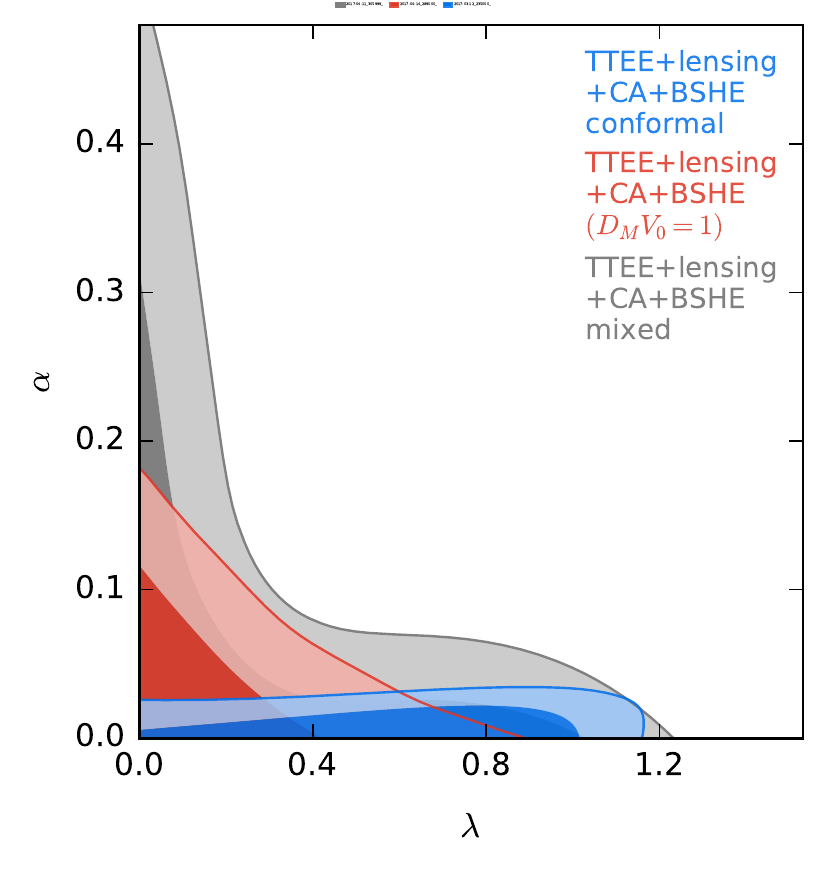}
\caption{A comparison of the marginalized two--dimensional constraints on the conformal coupling parameter $\alpha$, and the scalar field's potential parameter $\lambda$, in mixed models and the conformally coupled model. The mixed model with variable $D_M$ is denoted by mixed, whereas the other mixed model makes use of the relationship $D_M V_0=1$. In both mixed models, a constant disformal coupling $(\beta=0)$ was considered.}  
\label{fig:alpha_lambda_compare}
\end{figure}

Finally, we consider a mixed model having the same number of parameters as the conformal and disformal models considered in section \ref{sec:conformal_results} and section \ref{sec:disformal_results}, respectively. We report the MCMC analysis parameter constraints in the last column of Table \ref{table:mixed_Tab2}, in which we fix the constant disformal coupling parameter according to $D_M V_0=1$. As expected, we obtain tighter $95\%\;\mathrm{C.L.}$ upper bounds on the conformal coupling parameter $\alpha<0.1160$, as well as on the scalar field's exponent parameter $\lambda<0.5360$. We collect our marginalized constraints on $\alpha$ and $\lambda$ in Fig. \ref{fig:alpha_lambda_compare}, in which we compare the two--dimensional marginalized constraints inferred from the conformal model of section \ref{sec:conformal_results}, the mixed model with variable $D_M$ discussed in the first part of this section, along with the last mixed model which satisfies the relationship $D_M V_0=1$. From this figure, one can clearly observe that these models are all consistent with a null coupling between DM and DE, although the coupled DE models with a disformal coupling within the dark sector of the Universe still require further cosmological probes in order to properly determine the future prospects of these models.

\section{Conclusions}
\label{sec:conclusions}

The desire to better understand the puzzling dark sector of the Universe is one of the major driving forces that advances the field of precision cosmology even further. In this paper, we have considered a generic interacting DE model in which DE and DM were allowed to interact directly with each other. We showed that the cosmological imprints of this direct interaction between the dark sector constituents can be probed by current cosmological experiments. For the interaction between DM and DE, we considered the conformal and disformal couplings which are characterised by different cosmological signatures. We thus considered the conformal, disformal, and mixed models as separate cases of the generic coupled DE model. 

We confronted these coupled DE models with several combinations of data sets which are able to probe the early--time as well as the late--time cosmic history of the Universe. Specifically, we considered the \textit{Planck} 2015 temperature, polarization, and lensing likelihoods; BAO measurements, a SNIa sample, Hubble parameter measurements, local values of the Hubble constant, and cluster abundance measurements. The parameter posterior distributions together with their confidence limits were inferred via MCMC analyses.

In all coupled DE models, we found that the additional information from the cluster abundance data set and the CMB gravitational lensing likelihood improves the marginalized constraints on the coupling parameters. In general, we also noticed that the \textit{Planck} 2015 TTEE likelihood provides better marginalized constraints on the cosmological parameters, when compared with the inferred constraints from the CMB temperature likelihood. Also, the CA data set measurements which prefer relatively low values of $\sigma_8$, predominantly shift the marginalized constraints on $\tau_{\mathrm{reio}}$ and $A_s$ to lower values, with a proportionately smaller shift of $n_s$ to larger values. Moreover, we constrained the dimensionless age of the Universe in our coupled DE models, which we found to be close to unity without any significant changes from one model to another.

In the conformal model, the CMB likelihoods are able to constrain the model parameters quite well, since large values of the conformal coupling parameter have a significant impact on the CMB temperature power spectrum. With the additional information from the background data sets, we improved the constraints on $\lambda$, and marginally tighter constraints were placed on $\alpha$. Furthermore, the relatively high value of the local Hubble constant $H_0^{\mathrm{R}}$, gives rise to a significant peak in the marginalized posterior distribution of $\alpha$, although this is still found to be consistent with zero at $\sim2\sigma$. This complements Ref. \cite{Ade:2015rim}, in which a conformally coupled model with an inverse power--law potential was considered. However, with the additional cluster abundance measurements, the $95\%$ confidence level upper bounds on the coupling parameter are significantly lowered to $\alpha\lesssim0.03$. In our opinion, these tight limits on the conformal coupling between DM and DE diminishes the attractiveness of this model.

For the disformal model, we first considered a constant disformal coupling and then an exponential disformal coupling. Since a disformal coupling between DM and DE does not modify considerably the CMB temperature power spectrum, it was expected that the CMB likelihoods would not be able to constrain the model very well. Indeed, both the constant and exponential disformal coupling models were better constrained with the additional cluster abundance measurements which directly probe the characterised anomalous growth of perturbations. By being able to derive tight constraints on $\sigma_8$ from the information provided by the cluster abundance measurements, we were then able to place, for the first time, upper bounds on the disformal coupling parameters $D_M\lesssim0.3\,\mathrm{meV}^{-1}$ and $\beta\lesssim1.6$. Although the inferred constraints on the disformal model parameters are not as tight as those in the conformally coupled DE model, the disformal coupling is also consistent with a null coupling between the dark sector constituents. 

Finally, we considered the mixed conformally disformally coupled DE model. Similar to the previous models, the tightest constraints were obtained from the MCMC analyses which included the cluster abundance measurements. In this mixed model, significantly larger values of the conformal coupling parameter are allowed $(\alpha\lesssim0.33)$, in comparison with the derived upper bounds in the conformally coupled DE model. Since the disformal coupling parameter is not well constrained in the mixed model, we considered a mixed model which satisfies the relationship $D_M V_0=1$. In this case, we obtained the tightest marginalized constraint on the conformal coupling parameter of $\alpha\lesssim0.12$, which is still considerably larger than the $95\%$ confidence level upper bounds derived in the conformally coupled model. 

After confronting these coupled DE models with various cosmological probes, it would now be interesting to further constrain the DE couplings, particularly the disformal coupling, with 21--cm cosmology \cite{Kohri:2016bqx,Bull:2014rha,Maartens:2015mra} and gravitational waves \cite{Cai:2017yww,Wang:2017lgq,Caprini:2016qxs}, for instance. A better understanding of the non--linear evolution of perturbations in these models is also an important step in this direction.

\acknowledgments{The work of CvdB is supported by the Lancaster-Manchester-Sheffield Consortium for Fundamental Physics under STFC Grant No. ST/L000520/1.}

\bibliography{fullbib}

\end{document}